\documentclass[
superscriptaddress,
 amsmath,amssymb,
 aps,
 prl,
 twocolumn,
floatfix
longbibliography
]{revtex4-2}

\usepackage[colorlinks,linkcolor=blue,anchorcolor=blue,citecolor=blue,urlcolor=blue]{hyperref}
\usepackage[utf8]{inputenc}
\usepackage{graphicx}
\usepackage{threeparttable}
\usepackage{multirow,booktabs}
\usepackage{orcidlink}

\begin{document}

\date{\today}

\title{Strong evidence for \textsuperscript{9}N and  the limits of existence of atomic nuclei}

\def\WUPHYS{Department of Physics, Washington University, St. Louis, Missouri 63130, USA.}
\def\FRIB{Facility for Rare Isotope Beams, Michigan State University, East Lansing, Michigan 48824, USA.}
\def\PAMSU{Department of Physics \& Astronomy, Michigan State University, East Lansing, Michigan 48824, USA.}
\def\Fudan{Key Laboratory of Nuclear Physics and Ion-beam Application (MOE), Institute of Modern Physics, Fudan University, Shanghai 200433, China.}
\def\Shanghai{Shanghai Research Center for Theoretical Nuclear Physics, NSFC and Fudan University, Shanghai 200438, China.}
\def\MSUPHYS{Department of Physics and Astronomy, Michigan State University, East Lansing, Michigan 48824, USA.}
\def\MSUCHEM{Department of Chemistry, Michigan State University, East Lansing, Michigan 48824, USA.}
\def\WUCHEM{Department of Chemistry, Washington University, St. Louis, Missouri 63130, USA.}
\def\ANL{Physics Division, Argonne National Laboratory, Argonne, IL 60439, USA.}
\def\WesternM{Department of Physics, Western Michigan University, Kalamazoo, Michigan 49008, USA.}
\def\Stores{Department of Physics, University of Connecticut, Storrs, Connecticut 06269, USA.}
\def\Lanzhou{Institute of Modern Physics, Chinese Academy of Sciences, Lanzhou 730000, China.}

\author{R.J. Charity\orcidlink{0000-0003-3020-4998}} 
\affiliation{\WUCHEM}

\author {J. Wylie\orcidlink{0000-0002-0963-6190}}
\affiliation{\FRIB}
\affiliation{\PAMSU}

\author {S.M. Wang\orcidlink{0000-0002-8902-6842}}
\affiliation{\Fudan}
\affiliation{\Shanghai}

\author {T.B. Webb}
\affiliation{\WUPHYS}

\author {K.W. Brown}
\affiliation{\FRIB}

\author {G. Cerizza}
\affiliation{\FRIB}

\author {Z. Chajecki}
\affiliation{\WesternM}
\author {J.M. Elson}
\affiliation{\WUCHEM}
\author {J. Estee}
\affiliation{\FRIB}
\author {D.E.M Hoff}
\altaffiliation[Present address: ]{Lawrence Livermore National Laboratory, Livermore, CA 94550, USA.}
\affiliation{\WUCHEM}
\author {S.A. Kuvin}
\altaffiliation[Present address: ]{Los Alamos National Laboratory, Los Alamos, New Mexico 87545, USA.}

\affiliation{\Stores}
\author {W.G. Lynch}
\affiliation{\FRIB}
\affiliation{\PAMSU}
\author {J. Manfredi}

\altaffiliation[Present address: ]{Department of Engineering Physics, Air Force Institute of Technology, Wright-Patterson AFB, Ohio, 45433}
\affiliation{\FRIB}
\author {N. Michel}
\affiliation{\Lanzhou}
\author {D. G. McNeel}
\altaffiliation[Present address: ]{Los Alamos National Laboratory, Los Alamos, New Mexico 87545, USA.}
\affiliation{\Stores}
\author {P. Morfouace}
\affiliation{\FRIB}
\author {W. Nazarewicz\orcidlink{0000-0002-8084-7425}}
\affiliation{\FRIB}
\affiliation{\PAMSU}
\author {C.D. Pruitt}
\affiliation{\WUCHEM}
\author {C. Santamaria}
\affiliation{\FRIB}
\author {S. Sweany}
\affiliation{\FRIB}
\author {J. Smith}
\affiliation{\Stores}
\author {L.G. Sobotka\orcidlink{0000-0002-7883-7711}}
\affiliation{\WUCHEM}
\affiliation{\WUPHYS}
\author {M.B. Tsang}
\affiliation{\FRIB}
\author {A.H. Wuosmaa}
\affiliation{\Stores}

\begin{abstract}
The boundaries of the Chart of Nuclides contain   exotic isotopes 
that possess extreme proton-to-neutron asymmetries. Here we report on strong evidence of $^9$N, one of the most exotic proton-rich isotopes where more than one half of its constitute nucleons are unbound.    With seven protons and two neutrons, this extremely proton-rich system would represent the first-known example of a ground-state five-proton emitter. The invariant-mass spectrum of its decay products can be fit with two peaks whose energies are consistent with the theoretical predictions of an open-quantum-system approach, however we cannot rule out the possibility that only a single resonance-like peak is present in the spectrum. 
\end{abstract}

\maketitle
Nuclei with large imbalances between their constituent numbers of protons and neutrons can have exotic properties. The largest imbalances occur beyond the proton and neutron drip lines where  the nuclear ground states (g.s.) are unbound. Because of the  odd-even staggering of the drip lines induced by the nucleonic pairing, the shedding of unbound protons is usually terminated in an even-$Z$, particle-bound residue. Thus just beyond the proton drip line, one is likely to find single-proton emitters for odd-$Z$ isotopes and two-proton (2$p$) emitters for even-$Z$ isotopes \cite{Pfutzner:2012,Neufcourt2020p}. Even further removed, one finds 3$p$ and 4$p$ emitters.  Presently, 
\textsuperscript{7}B, \textsuperscript{13}F, \textsuperscript{17}Na, and \textsuperscript{31}K are the known 3$p$ emitters \cite{Charity:2011,Charity:2021,Brown:2017,Kostyleva:2019} and  \textsuperscript{8}C and \textsuperscript{18}Mg have been observed to decay by emission of four protons \cite{Charity:2010, Jin:2021}.

Moving outward past the drip lines, the decay widths of the low-lying states increase, eventually melting into an unresolvable continuum as their lifetimes become commensurate with typical reaction and single-particle timescales.
Here, the very notion of the {\it nuclear state} becomes questionable as the timescales are too short to talk about the existence of a  nucleus. Indeed as discussed in Ref.~\cite{Thoennessen2004}, if a collection of nucleons survives for less than about $10^{-22}$\,s, it should not be  considered a nucleus.
In this regime, the decay properties manifest themselves as {\it scattering features}  rather than well-defined resonances.  The maximum decay width at the boundary for $A\approx8$,  based on  single-particle timescales, is of order $\Gamma \approx$ 3.5~MeV \cite{Fossez:2016}.

\begin{figure}[!htb]
\includegraphics[width=0.7\linewidth]{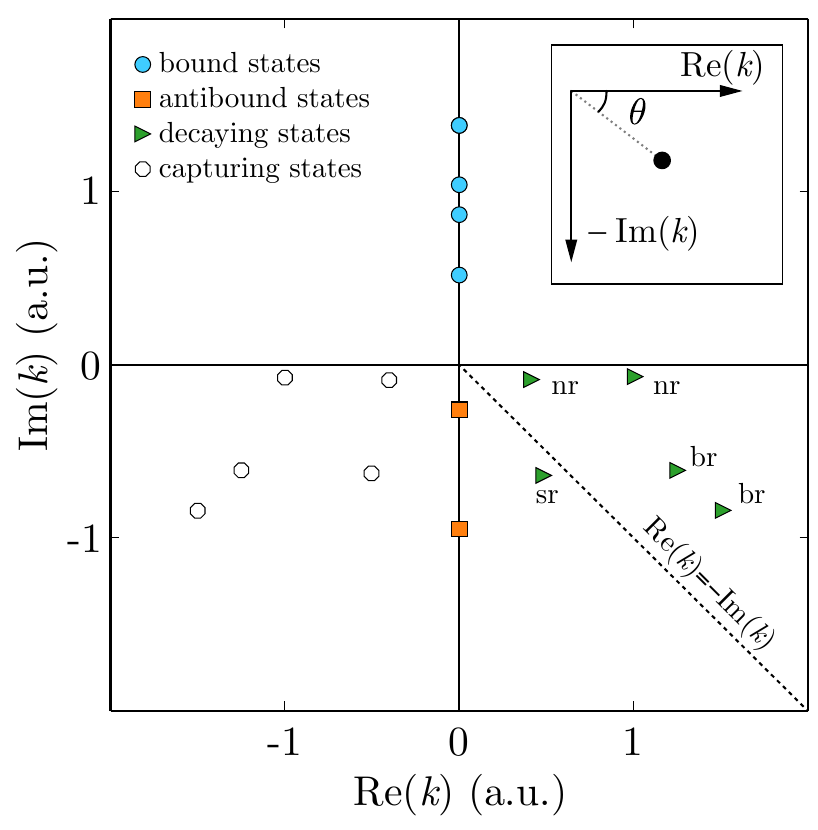}
\caption{Classification of resonant states based on the location of the scattering-matrix poles in the complex-$k$ plane. The same arbitrary units (a.u.) are used on both axes. Bound, antibound, decaying, and capturing resonant states are marked, as well as narrow resonances (nr), broad resonant states (br) and subthreshold resonant states (sr). The distribution of poles is symmetric with respect to the imaginary $k$-axis because of time reversal symmetry; thus, capturing states are presented as the time-reversed decaying states. The dashed  $-45^\circ$ line 
separates decaying resonant states from subthreshold poles. The inset shows  the polar angle $\theta$ of the resonant pole in the complex momentum plane.
} 
\label{fig:poles}
\end{figure}

The $4p$ emitter \textsuperscript{8}C has one half of its nucleons in the continuum and the remainder constitutes an $\alpha$-cluster. The nucleus $^9$N is even more extreme with an additional proton in the continuum. In nuclei such as $^8$C and $^9$N, we might anticipate difficulty in classifying nuclear properties in terms  of localized nuclear states. On the other hand, new insights can be gained from such unbound clusters of nucleons. 
In this Letter, we make a careful distinction between {\it resonances} (sharp peaks in the experimental scattering cross section) and {\it resonant states} (complex-momentum poles of the scattering matrix). For the theoretical classification of resonant states, see Fig.\,\ref{fig:poles}.  The decaying resonant states in the fourth quadrant of the complex-$k$ plane,  lying close to the real $k$-axis and having a real energy Re$(E)>0$ and a width $\Gamma$=$-$2Im$(E)>0$, can be interpreted as narrow resonances seen in experiments. The poles with Re$(E)<0$ and $\Gamma>0$, can be associated with subthreshold resonant states ($\theta>45^\circ$). The antibound states have Re$(E)<0$ and $\Gamma=0$.

Unlike the g.s. of \textsuperscript{8}C, the dineutron is not a genuine resonance 
(energy $E>0$, decay width $\Gamma>0$), but an  antibound state (or a scattering feature) \cite{Ohanian1974,Babenko:2013} located on the second Riemann energy sheet. Here the attractive interaction between the two neutrons is just insufficient to produce a bound state, but its nearly-bound nature is manifested by enhanced $n$+$n$ scattering just above threshold and by strong final-state effects.  The diproton is a subthreshold resonant state ($E<0, \Gamma>0$) \cite{Kok:1980,Mukhamedzhanov2010} and located below the $-45^\circ$ line in the complex-$k$ plane (see Fig.\,\ref{fig:poles}). It is formally neither an antibound state nor a resonance (though closer to the latter), and again manifests itself by enhanced scattering strength and final-state effects. While there have been some recent suggestions of a tetra-neutron resonance-like structure, what was observed \cite{Kisamori:2016,Duer:2022} may instead be a final-state effect  \cite{Deltuva:2018,Higgins:2020}. Similar situations might occur in the g.s. of $^9$N and $^9$He as presented in this work. 
 
The nucleus \textsuperscript{9}N has three neutrons less than  the lightest particle-bound nitrogen isotope \textsuperscript{12}N and one more proton than \textsuperscript{8}C into which it decays. The neighboring isotope \textsuperscript{10}N has only been observed in three studies \cite{Lepine-Szily:2002,Hooker:2017, Charity:2021b}. It has low-lying states which are single-proton  resonances although their structure is not well established. Some indication as to the structure of \textsuperscript{9}N can be gleaned from its mirror partner \textsuperscript{9}He for which low-energy resonance-like states decay by the $n$+$^{8}$He channel.

Particular interest in \textsuperscript{9}He is due to the parity inversion of the ground-state spin for odd $N=7$ isotones with large neutron excesses when the second $s_{1/2}$ neutron single-particle orbital intrudes into the $p$ shell \cite{Aumann:2000}. Most studies agree that there is a 1/2$^-$ resonance around 1.2~MeV above the $n$+\textsuperscript{8}He threshold (see Ref. \cite{Kalanee:2013}), but there is less agreement about its width \cite{Bohlen:1988,Kalanee:2013,Votaw:2020}. A number of studies find some 1/2$^+$ strength below this resonance \cite{Chen:2001,Golovkov:2007,Johansson:2010,Kalanee:2013,Votaw:2020} although this strength in some studies in not sufficient to justify a notion of a  state \cite{Johansson:2010,Votaw:2020,Vorabbi:2018}.

{\it Experiment.---} An $E/A$=69.5\,MeV secondary beam of \textsuperscript{13}O (4$\times$10$^5$ pps, purity 80\%)  was produced from the Coupled Cyclotron Facility at the National Superconducting Cyclotron Laboratory at Michigan State University.  Charged particles created in the interaction with a 1\,mm-thick $^9$Be target  were detected in the High Resolution Array (HiRA) \cite{Wallace:2007} consisting of 14 $E$-$\Delta E$ telescopes covering scattering angles from 2.1$^{\circ}$ to 12.4$^\circ$.
The location of the HiRA telescopes was chosen based on Monte Carlo simulations which optimised the detection of low-lying resonances produced in projectile-fragmentation reactions. More details of the experiment and the simulations can be found in 
the Supplemental Material\,\cite{sup}. States in \textsuperscript{9}N  were produced in fragmentation reactions where 3 neutrons and 1 proton in total are removed from the projectile and the resonance is identified using the invariant-mass technique.  Data from this experiment pertaining to the first observation of  $^{11}$O  \cite{Webb:2019a,Webb:2020} and $^{13}$F \cite{Charity:2021} as well as other 
previously-known  isotopes  \cite{Webb:2019,Charity:2021a,Charity:2021b} have already been published.

\begin{figure}[!htb]
\centering
\includegraphics[width=1.\linewidth]{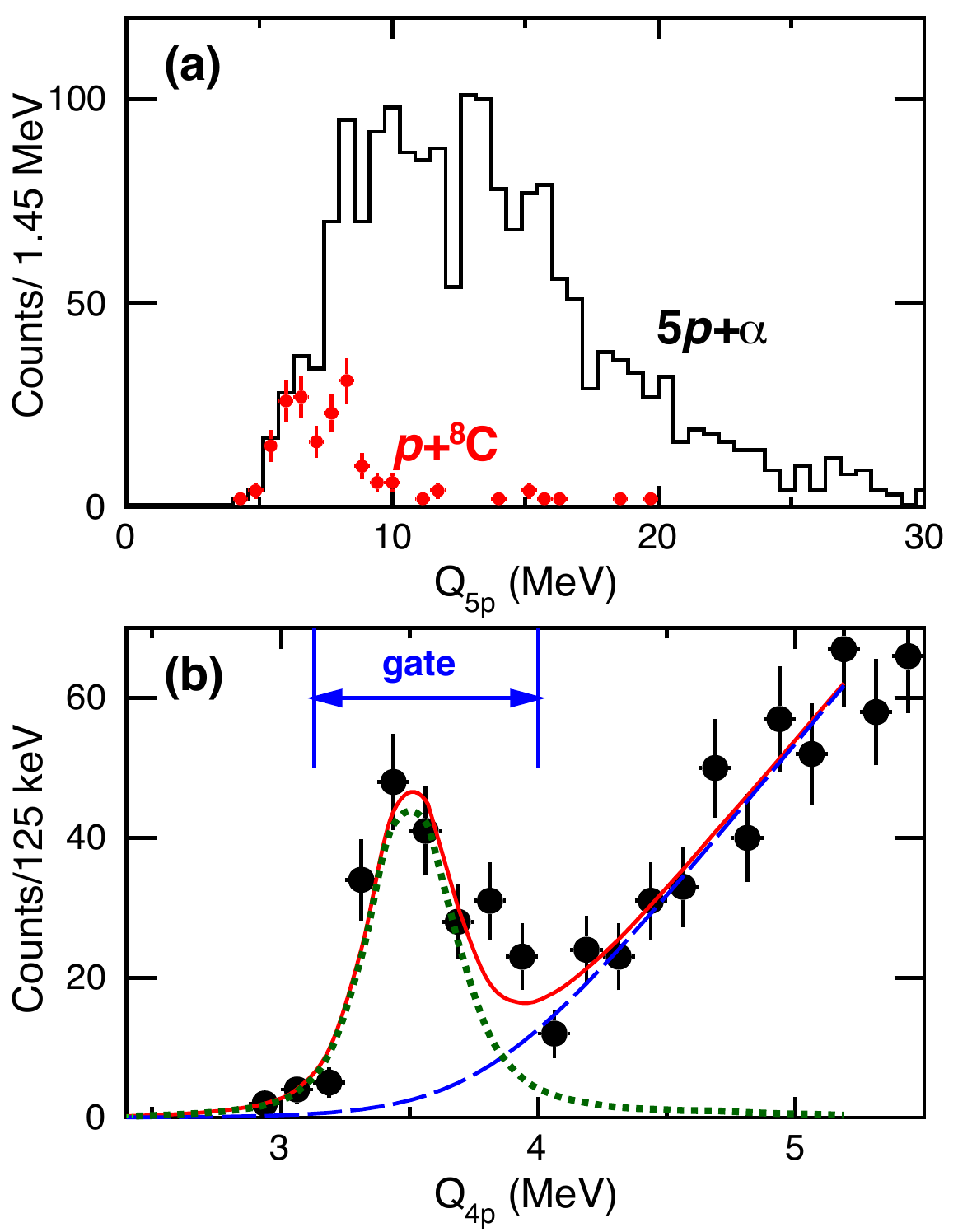}
\caption{Invariant-mass spectra for \textsuperscript{9}N and \textsuperscript{8}C. (a) Histogram shows the distribution of the decay energy $Q_{5p}$ obtained from 5$p$+$\alpha$ events with the invariant-mass method. The data points show the distribution where an $^8$C(g.s.) intermediate state was identified. (b) The data points show the distribution of $^8$C decay energy ($Q_{4p}$) from the five 4$p$+$\alpha$ subevents in each 5$p$+$\alpha$ event. The solid red curve shows a fit to this distribution using the experimental $^8$C lineshape obtained from detected 4$p$+$\alpha$ events (dotted green curve) plus a smooth background (dashed blue curve). The gate used to select $p$+$^8$C(g.s.) events is indicated. }
\label{fig:N9C8}
\end{figure}

{\it Invariant-Mass Spectra.---} 
The decay-energy ($Q_{5p}$) distribution for all detected 5$p$+$\alpha$ events [Fig.\,\ref{fig:N9C8}(a)] is quite wide and contains no prominent ``narrow'' peaks but there can be contributions from the decay of very wide \textsuperscript{9}N states and non-resonant breakup. These events are produced in projectile-fragmentation reactions as their center-of-mass velocities are close to the beam value \cite{Charity:2023}. Given that the lowest-energy states in the mirror system \textsuperscript{9}He decay to the $n$+\textsuperscript{8}He(g.s.) channel, it is thus interesting to look at the mirror $p$+\textsuperscript{8}C(g.s.) channel. For each 5$p$+$\alpha$ event, we can remove one of the five protons in turn to create the five possible 4$p$+$\alpha$ subevents. The distribution of $^{8}$C decay energy ($Q_{4p}$) from these subevents [ Fig.\,\ref{fig:N9C8}(b)] has a peak corresponding to the g.s. of \textsuperscript{8}C at the expected value of $Q_{4p}$. For comparison, the dotted green curve shows the shape of the \textsuperscript{8}C peak obtained from fitting the more numerous 4$p$+$\alpha$ events (see Supplemental Material)  

 Possible $^9$N$\rightarrow p$+\textsuperscript{8}C(g.s.) decay events were selected by requiring that at least one of the 4$p$+$\alpha$ subevents lies in the gate around the ground-state peak  [Fig.\,\ref{fig:N9C8}(b)]. Based on the presented fit to the ground-state peak in this figure, 20\% of the 4$p$+$\alpha$ subevents events in this gate are associated with the background component. However the gate on $p$+\textsuperscript{8}C(g.s.) is cleaner than this as some of the events contribute to both the peak and this background through their different subevents. Indeed 11\% of selected events have two subevents in this gate which accounts for 1/2 of the background contamination. These numerical partitions are confirmed by our simulations.

 The final distribution, with the $^8$C(g.s.) gate, is largely restricted to $Q_{5p}<$10~MeV [data points Fig.~\ref{fig:N9C8}(a)] and appears to have two peaks. Hence it may be a doublet rather than a singlet although the statistics are marginal for this distinction.

{\it Theoretical models.---} Since  continuum effects are strong  for both $^8$C and $^9$N, we used the complex-energy Gamow Shell Model (GSM) to predict the location of the nuclear states of interest as it has been used to study many weakly-bound and unbound systems \cite{Michel:2002,Michel2009,Michel:2021} including the g.s. of $^8$C \cite{Wylie:2021}.
GSM differs from the traditional closed-quantum-system nuclear shell model as it allows for bound, scattering, and Gamow resonant states with outgoing asymptotic behavior  to be treated on equal footing by implementing the Berggren basis \cite{Berggren:1968}.

Calculations for $^8$C and $^9$N were performed by assuming an $\alpha$ core surrounded by  four and five valence protons, respectively. We used the same valence-space Hamiltonian as described in previous GSM studies \cite{Jaganathen:2017,Mao:2020,Wylie:2021} with the parameters given in  the Supplemental Material. 
The maximum number of particles allowed in the continuum is 2 for the results presented; it has been checked that more particles in the continuum barely change the conclusion for $A=8$ system.
The predicted $Q_{4p}$ values for the g.s. and the 2$^+$ first excited state of $^8$C, shown in Fig.~\ref{fig:level} with the relevant levels for their decay, are around 3.7 and 8.3\,MeV respectively (see Supplemental Material; see also Ref.~\cite{Myo:2021} for complex-scaling predictions of $^8$C). 

\begin{figure}[!htb]
\centering
\includegraphics[width=1.0\linewidth]{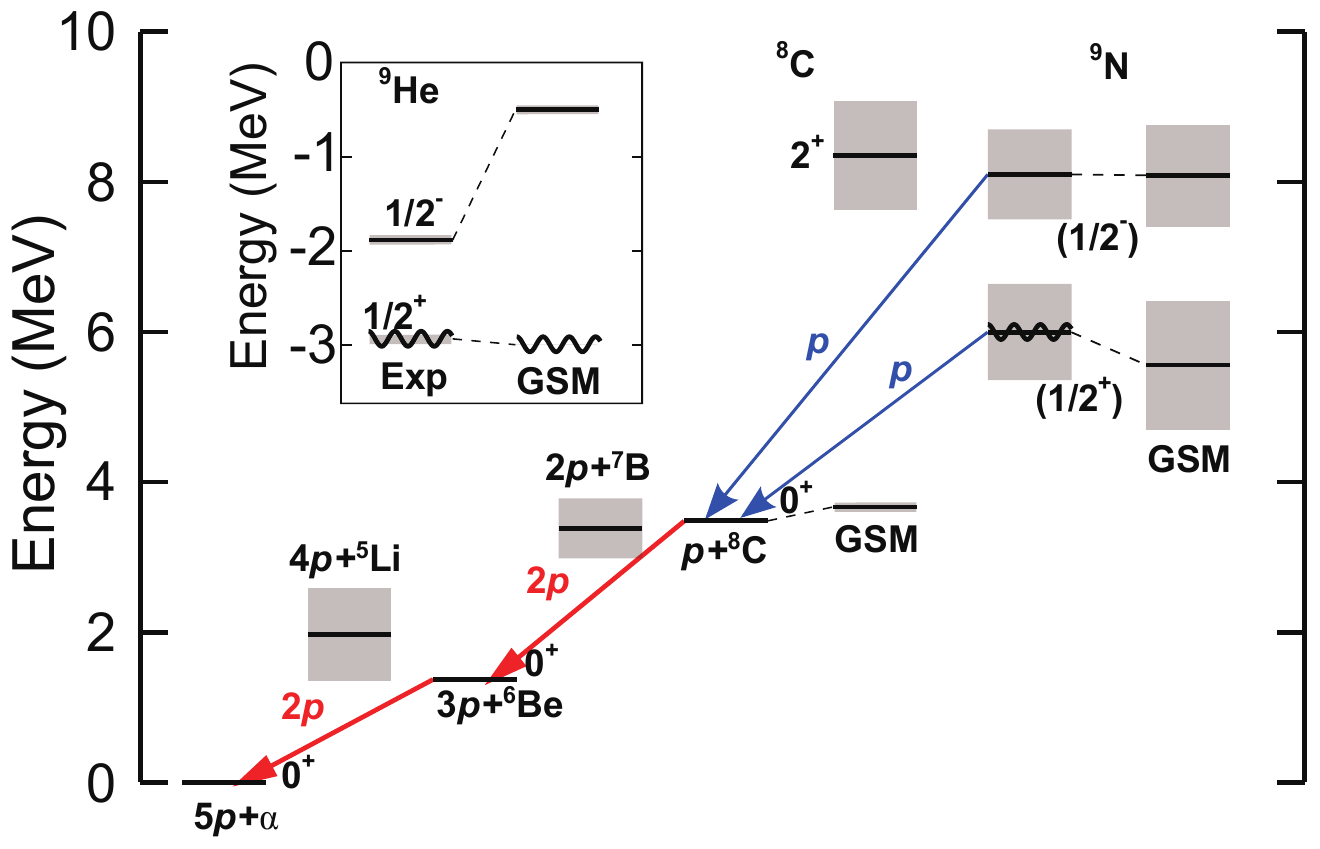}
\caption{ Level diagrams of $^8$C and $^9$N obtained experimentally and calculated with the GSM. Energies are  given relative to the $^4$He threshold. The level diagram of  $^9$He, the mirror partner of $^9$N, is shown in the inset. The $1/2^+$ antibound state in $^9$He is shown with a wavy line to indicate its status more as a scattering feature rather than a real state. The proposed $1/2^+$ state in $^9$N is shown with both straight and wavy lines to indicate the uncertainty as to its nature (a resonance  or scattering feature) while the GSM interprets it  as a broad resonant state.  } 
    \label{fig:level}
\end{figure}


The properties of the  low-lying states in $^9$N were crosschecked by the Gamow-Coupled-Channel (GCC) method \cite{Wang:2019,Wang:2021}, which is a complex framework utilizing the  Berggren basis. Based on large proton-decaying spectroscopic factors (${\cal S}^2_{s_{1/2}}$ = 0.87 and ${\cal S}^2_{p_{1/2}}$ = 0.80 according to GSM calculations), $^9$N can be described as a $^8$C+$p$ two-body system.

{\it Discussion.---} 
While we have cleanly selected $p$+$^8$C(g.s.) events, the extra proton beyond the $^8$C decay products can result from either $^9$N resonant decay or alternatively is produced directly in the first step of the projectile-fragmentation reaction. To form $^9$N,  one proton and three neutrons need to be removed from the projectile and thus one prompt proton would be expected unless it is bound in a $d$ or $t$ cluster with some of the promptly removed neutrons. Also, there can be contributions from  events where the projectile promptly breaks up into the channel 2$p$+3$n$+$^8$C(g.s).
While most of the promptly emitted protons are expected at larger angles than sampled with HiRA, we do expect some background from these prompt particles \cite{Charity:2023}.

The background from all these sources of prompt protons 
 has been extensively studied in Ref.~\cite{Charity:2023} for this and other resonant-decay channels associated with this projectile. The $Q_{5p}$ dependence of this background can be obtained from mixing $p$ and $^8$C decay products from different $p$+$^8$C(g.s.) events appropriately weighted to include correlations between them. However, this predicted $Q_{5p}$ distribution is too wide to explain all the observed events  and thus substantial  contributions from $^9$N resonances are required \cite{Charity:2023}.  Indeed the possibility that statistical fluctuations in this background can explain the observed spectrum can be ruled out with a significance of roughly 5$\sigma$, i.e. at the discovery threshold for a new resonance (see Supplemental Material). 

The GSM predicts  1/2$^+$ and 1/2$^-$ states in the region below $Q_{5p}$=10~MeV where the $p$+$^{8}$C(g.s.) yield is strongest. The predicted ground-state has the same parity-inversion as observed for $^{11}$Be \cite{Aumann:2000}. With two  states and the magnitude of the background to vary, there are 7 fit parameters which is about the same as the number of data points in the peak region. Thus the experimental statistics are presently not sufficient to fully constrain the location of both resonance poles ($Q_{1p}-i\Gamma$/2) and determined their nature 
(genuine or sub-threshold resonances) from Fig.~\ref{fig:poles}. In the following, in order to show some possible interpretations of the data,  we present some constrained fits which avoid overfitting.

\begin{figure}[!htb]
\centering
\includegraphics[width=1.\linewidth]{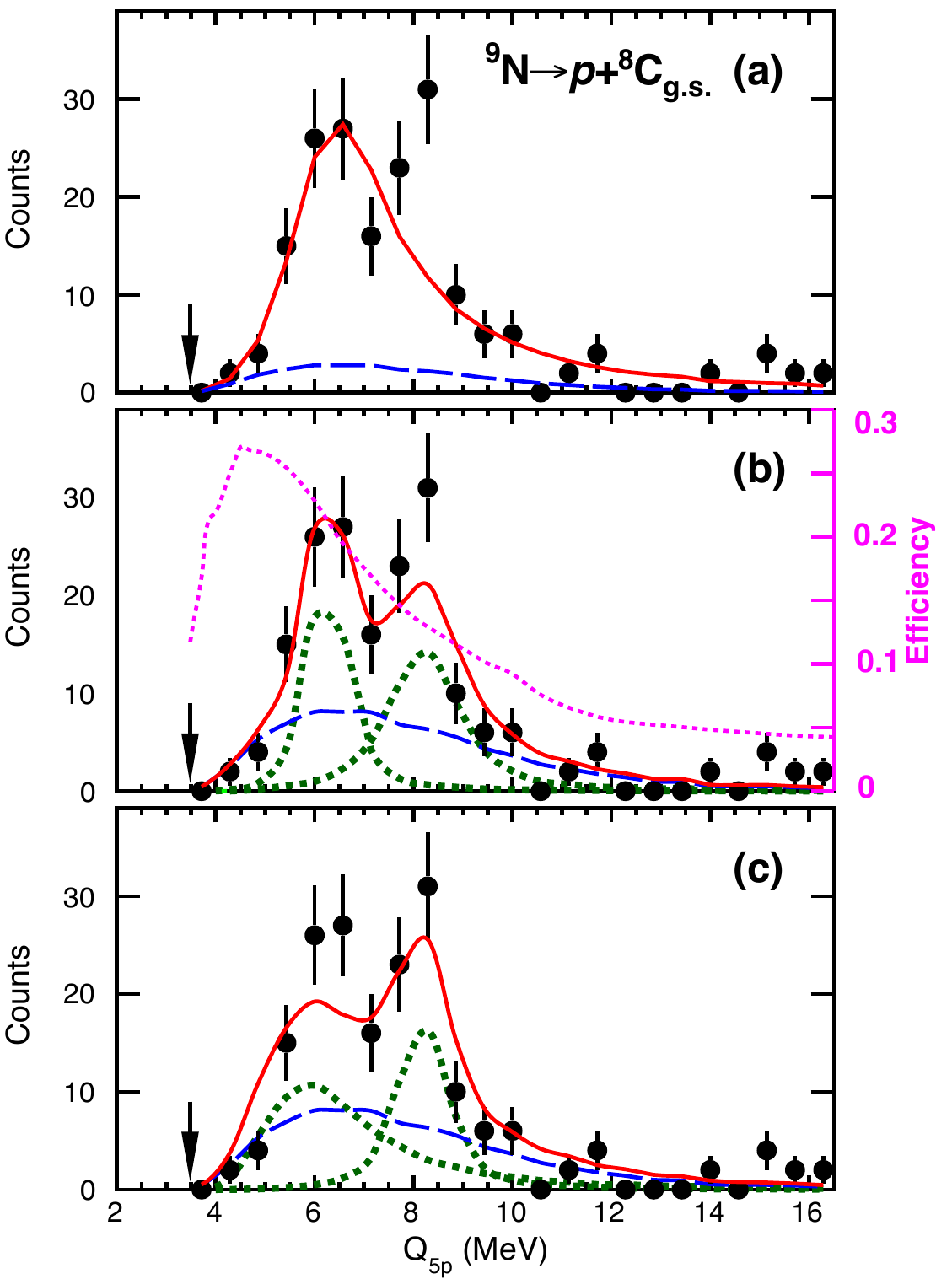}
\caption{Fits (solid red curves) to the selected $p$+\textsuperscript{8}C(g.s.) distribution. 
(a)  Fit with a single-peak parameterized by an $\ell=1$ $R$-matrix lineshape. (b-c) Two peak fits where the contributions from both levels are given by the dotted green curves. 
The fit in (b) was using a $p$+$^8$C(g.s.) $R$-matrix lineshape where energy and width of the upper peak are set to the GSM predictions for the $J^\pi$=1/2$^-$ state of \textsuperscript{9}N.  The dependence of the detection efficiency as a function of $Q_{5p}$ as determined in our Monte Carlo simulations 
(see Supplemental Material) is shown by the dashed-magenta curve. The fit in (c) was obtained where the fixed width of the 1/2$^-$ was halved from its value in (b) and the 1/2$^+$ strength is described by a sub-threshold resonant state using GCC lineshape. In all panels, the $p$+\textsuperscript{8}C(g.s.) threshold is indicated by an arrow and the fitted background from nonresonant protons is shown by the dashed blue curves.} 
\label{fig:fit}
\end{figure}

At one extreme, we can assume just a single peak plus background. Using $R$-matrix line shapes \cite{Lane:1958} for the  $p$+$^8$C(g.s.) channel, the qualities of the fits were very similar for both $s$ and $p$ resonances. The resonance  pole
 for a single $s$-wave fit is at $Q_{1p}$ = 1.22(16)~MeV [$Q_{5p}$=4.70(16)] and $\Gamma$=2.59(23) MeV.
Such a state is approaching the diffuse borderland between a broad resonance and a scattering feature. However this solution is disfavored as the magnitude of the background has been reduced to zero to achieve the best fit. The result is not much better for a single $p$-wave resonance [Fig.~\ref{fig:fit}(a)] with the resonance pole given by $Q_{1p}$=2.01(16)~MeV [$Q_{5p}$=5.49(16)~MeV] and $\Gamma$=2.28(23)~MeV. In this case, the fitted background (dash blue curve) contributes only 14\% of the total yield. As other resonant channels from this $^{13}$O-induced reaction had backgrounds of magnitude 30-54\%  \cite{Charity:2023}, these single-peak solutions are disfavored.

Figure~\ref{fig:fit}(b) shows an example of a possible
two-peak fit assuming real $p$+$^8$C(g.s.) resonances described by $R$-matrix line shapes.  Here, the $R$-matrix parameters of the 1/2$^-$ peak are fixed so that its pole is consistent with the GSM prediction. With this constraint, the fitted values for the 1/2$^+$ strength are $Q_{1p}$=2.75(21)~MeV [$Q_{5p}$=6.23(21)~MeV] and $\Gamma$=0.58(44)~MeV which can be compared to the GSM predictions of $Q_{5p}$=5.56~MeV and $\Gamma$=1.74~MeV. 
The fitted background from prompt protons (dashed blue curve) now explains the observed yields below $Q_{5p}$=5~MeV and the high-energy tail above 10~MeV. This background accounts for $\approx$40\% of the yield which is similar to  values of 30-54\% obtained in fitting the $^{10,11}$N$\rightarrow p$+$^{9,10}$C invariant-mass distributions \cite{Charity:2023}.
While this fitted width of the 1/2$^+$ state is smaller than the GSM prediction, it can be increased by decreasing the  fixed width of the 1/2$^-$ state.  For example, Fig.~\ref{fig:fit}(c) shows a fit where the intrinsic width of the 1/2$^-$ state has been halved, but now the 1/2$^+$ strength is described by  the GCC lineshape as a subthreshold resonant state.

To provide insights into the nature of the $1/2^+$ state in $^9$N, GSM calculations were performed for its mirror partner, $^9$He, using the same Hamiltonian parameters.
Figure\,\ref{fig:level} shows that the experimental spectrum is reasonably reproduced, although the prediction for the 1/2$^-$ state is somewhat too high in energy. 
The previous studies using the Berggren basis have indicated that the $1/2^+$ state in $^9$He is a resonance \cite{Jaganathen:2017,Fossez:2018}, in which some bound poles were introduced in the basis to stabilize their results (see Supplemental Material for details). However, when the continuum effect is properly taken into account in the current work with a deformed scattering contour \cite{Michel2006a}, one can generate an  antibound $1/2^+$ pole in $^9$He, an approach yielding the best agreement with experimental data.

Although it is not possible for antibound poles to emerge in proton-rich nuclei due to the Coulomb  interaction \cite{Wang:2019}, we followed the same procedure as in $^9$He to determine if the $1/2^+$ state in $^9$N is a broad resonant state or subthreshold resonant state.
The predicted $1/2^+$ state has an energy of $E=2.08\;$MeV above the $^8$C$+p$ threshold and $\Gamma=1.74\;$MeV which indicates a  broad resonant state as defined in Fig.~\ref{fig:poles}. It must be noted, however, that the $1/2^+$ state in GSM is very fragile with respect to changes of the Berggren basis due to the truncations (particle-hole, discretization of the continuum) employed (see Supplemental Material for details), so we cannot with certainty rule out a subthreshold resonant state.

Both $^8$C and $^9$N shed their excess protons by sequential steps of either single-proton or prompt $2p$ decays. The $4p$-emitter $^{18}$Mg also decays in this manner \cite{Jin:2021}, suggesting this behavior is typical for isotopes at the limits of existence. Our analysis suggests that some observed structures,  sometimes interpreted in terms of nuclear states, should be  viewed rather as fleeting features lying outside the Chart of Nuclides.

{\it Conclusions.---} We have found strong evidence for the exotic nuclide $^9$N produced in the fragmentation of a $^{13}$O beam.  The invariant-mass spectrum of detected 5$p$+$\alpha$ events, each containing an  $^8$C(g.s.) intermediate state, contains a structure  which cannot be explained  at the $\approx$5$\sigma$ level by statistical fluctuations of the expected background from other coincident protons liberated in the fragmentation event.  This observed structure can be interpreted as two $^9$N peaks, although a single-peak solution cannot be totally discounted. The $^9$N state(s) would be first known case(s) of  5$p$ emission from ground-state and low-lying resonances. 

This nuclide has also been studied theoretically in the Gamow Shell Model where the important effects of the continuum are included. The predicted locations  of the 1/2$^+$ and 1/2$^-$ resonant states are in excellent agreement with the location of the observed structure giving further evidence for the preferred two-peak solution. In the model, the 1/2$^+$ ground state of $^9$N is the mirror of an antibound state in $^9$He, and is most likely a broad resonant state rather than a subthreshold resonant state. However the latter cannot be completely ruled out in both the experiment and the theory.

 \begin{acknowledgments}
 This material is based upon work supported by the U.S.
Department of Energy, Office of Science, Office of Nuclear
Physics under Awards No. DE-FG02-87ER-40316,
No. DE-FG02-04ER-41320, No. DE-SC0014552, DOE-DE-SC0013365,
DE-SC0023175,
the National Science Foundation under Grant No. PHY-
156556, the National Key Research and Development Program (MOST 2022YFA1602303), and the National Natural Science Foundation of China under Grant Nos. 12175281, 12147101. J. M. was supported by a Department of Energy
National Nuclear Security Administration Steward Science
Graduate Fellowship under cooperative agreement No. DE-
NA0002135.
  \end{acknowledgments}



\begin{thebibliography}{60}%
\makeatletter
\providecommand \@ifxundefined [1]{%
 \@ifx{#1\undefined}
}%
\providecommand \@ifnum [1]{%
 \ifnum #1\expandafter \@firstoftwo
 \else \expandafter \@secondoftwo
 \fi
}%
\providecommand \@ifx [1]{%
 \ifx #1\expandafter \@firstoftwo
 \else \expandafter \@secondoftwo
 \fi
}%
\providecommand \natexlab [1]{#1}%
\providecommand \enquote  [1]{``#1''}%
\providecommand \bibnamefont  [1]{#1}%
\providecommand \bibfnamefont [1]{#1}%
\providecommand \citenamefont [1]{#1}%
\providecommand \href@noop [0]{\@secondoftwo}%
\providecommand \href [0]{\begingroup \@sanitize@url \@href}%
\providecommand \@href[1]{\@@startlink{#1}\@@href}%
\providecommand \@@href[1]{\endgroup#1\@@endlink}%
\providecommand \@sanitize@url [0]{\catcode `\\12\catcode `\$12\catcode
  `\&12\catcode `\#12\catcode `\^12\catcode `\_12\catcode `\%12\relax}%
\providecommand \@@startlink[1]{}%
\providecommand \@@endlink[0]{}%
\providecommand \url  [0]{\begingroup\@sanitize@url \@url }%
\providecommand \@url [1]{\endgroup\@href {#1}{\urlprefix }}%
\providecommand \urlprefix  [0]{URL }%
\providecommand \Eprint [0]{\href }%
\providecommand \doibase [0]{https://doi.org/}%
\providecommand \selectlanguage [0]{\@gobble}%
\providecommand \bibinfo  [0]{\@secondoftwo}%
\providecommand \bibfield  [0]{\@secondoftwo}%
\providecommand \translation [1]{[#1]}%
\providecommand \BibitemOpen [0]{}%
\providecommand \bibitemStop [0]{}%
\providecommand \bibitemNoStop [0]{.\EOS\space}%
\providecommand \EOS [0]{\spacefactor3000\relax}%
\providecommand \BibitemShut  [1]{\csname bibitem#1\endcsname}%
\let\auto@bib@innerbib\@empty
\bibitem [{\citenamefont {Pf\"utzner}\ \emph {et~al.}(2012)\citenamefont
  {Pf\"utzner}, \citenamefont {Karny}, \citenamefont {Grigorenko},\ and\
  \citenamefont {Riisager}}]{Pfutzner:2012}%
  \BibitemOpen
  \bibfield  {author} {\bibinfo {author} {\bibfnamefont {M.}~\bibnamefont
  {Pf\"utzner}}, \bibinfo {author} {\bibfnamefont {M.}~\bibnamefont {Karny}},
  \bibinfo {author} {\bibfnamefont {L.~V.}\ \bibnamefont {Grigorenko}},\ and\
  \bibinfo {author} {\bibfnamefont {K.}~\bibnamefont {Riisager}},\ }\bibfield
  {title} {\bibinfo {title} {Radioactive decays at limits of nuclear
  stability},\ }\href {https://doi.org/10.1103/RevModPhys.84.567} {\bibfield
  {journal} {\bibinfo  {journal} {Rev. Mod. Phys.}\ }\textbf {\bibinfo {volume}
  {84}},\ \bibinfo {pages} {567} (\bibinfo {year} {2012})}\BibitemShut
  {NoStop}%
\bibitem [{\citenamefont {Neufcourt}\ \emph {et~al.}(2020)\citenamefont
  {Neufcourt}, \citenamefont {Cao}, \citenamefont {Giuliani}, \citenamefont
  {Nazarewicz}, \citenamefont {Olsen},\ and\ \citenamefont
  {Tarasov}}]{Neufcourt2020p}%
  \BibitemOpen
  \bibfield  {author} {\bibinfo {author} {\bibfnamefont {L.}~\bibnamefont
  {Neufcourt}}, \bibinfo {author} {\bibfnamefont {Y.}~\bibnamefont {Cao}},
  \bibinfo {author} {\bibfnamefont {S.}~\bibnamefont {Giuliani}}, \bibinfo
  {author} {\bibfnamefont {W.}~\bibnamefont {Nazarewicz}}, \bibinfo {author}
  {\bibfnamefont {E.}~\bibnamefont {Olsen}},\ and\ \bibinfo {author}
  {\bibfnamefont {O.~B.}\ \bibnamefont {Tarasov}},\ }\bibfield  {title}
  {\bibinfo {title} {Beyond the proton drip line: Bayesian analysis of
  proton-emitting nuclei},\ }\href
  {https://doi.org/10.1103/PhysRevC.101.014319} {\bibfield  {journal} {\bibinfo
   {journal} {Phys. Rev. C}\ }\textbf {\bibinfo {volume} {101}},\ \bibinfo
  {pages} {014319} (\bibinfo {year} {2020})}\BibitemShut {NoStop}%
\bibitem [{\citenamefont {Charity}\ \emph {et~al.}(2011)\citenamefont
  {Charity}, \citenamefont {Elson}, \citenamefont {Manfredi}, \citenamefont
  {Shane}, \citenamefont {Sobotka}, \citenamefont {Brown}, \citenamefont
  {Chajecki}, \citenamefont {Coupland}, \citenamefont {Iwasaki}, \citenamefont
  {Kilburn}, \citenamefont {Lee}, \citenamefont {Lynch}, \citenamefont
  {Sanetullaev}, \citenamefont {Tsang}, \citenamefont {Winkelbauer},
  \citenamefont {Youngs}, \citenamefont {Marley}, \citenamefont {Shetty},
  \citenamefont {Wuosmaa}, \citenamefont {Ghosh},\ and\ \citenamefont
  {Howard}}]{Charity:2011}%
  \BibitemOpen
  \bibfield  {author} {\bibinfo {author} {\bibfnamefont {R.~J.}\ \bibnamefont
  {Charity}}, \bibinfo {author} {\bibfnamefont {J.~M.}\ \bibnamefont {Elson}},
  \bibinfo {author} {\bibfnamefont {J.}~\bibnamefont {Manfredi}}, \bibinfo
  {author} {\bibfnamefont {R.}~\bibnamefont {Shane}}, \bibinfo {author}
  {\bibfnamefont {L.~G.}\ \bibnamefont {Sobotka}}, \bibinfo {author}
  {\bibfnamefont {B.~A.}\ \bibnamefont {Brown}}, \bibinfo {author}
  {\bibfnamefont {Z.}~\bibnamefont {Chajecki}}, \bibinfo {author}
  {\bibfnamefont {D.}~\bibnamefont {Coupland}}, \bibinfo {author}
  {\bibfnamefont {H.}~\bibnamefont {Iwasaki}}, \bibinfo {author} {\bibfnamefont
  {M.}~\bibnamefont {Kilburn}}, \bibinfo {author} {\bibfnamefont
  {J.}~\bibnamefont {Lee}}, \bibinfo {author} {\bibfnamefont {W.~G.}\
  \bibnamefont {Lynch}}, \bibinfo {author} {\bibfnamefont {A.}~\bibnamefont
  {Sanetullaev}}, \bibinfo {author} {\bibfnamefont {M.~B.}\ \bibnamefont
  {Tsang}}, \bibinfo {author} {\bibfnamefont {J.}~\bibnamefont {Winkelbauer}},
  \bibinfo {author} {\bibfnamefont {M.}~\bibnamefont {Youngs}}, \bibinfo
  {author} {\bibfnamefont {S.~T.}\ \bibnamefont {Marley}}, \bibinfo {author}
  {\bibfnamefont {D.~V.}\ \bibnamefont {Shetty}}, \bibinfo {author}
  {\bibfnamefont {A.~H.}\ \bibnamefont {Wuosmaa}}, \bibinfo {author}
  {\bibfnamefont {T.~K.}\ \bibnamefont {Ghosh}},\ and\ \bibinfo {author}
  {\bibfnamefont {M.~E.}\ \bibnamefont {Howard}},\ }\bibfield  {title}
  {\bibinfo {title} {Investigations of three-, four-, and five-particle decay
  channels of levels in light nuclei created using a $^{9}\mathrm{C}$ beam},\
  }\href {https://doi.org/10.1103/PhysRevC.84.014320} {\bibfield  {journal}
  {\bibinfo  {journal} {Phys. Rev. C}\ }\textbf {\bibinfo {volume} {84}},\
  \bibinfo {pages} {014320} (\bibinfo {year} {2011})}\BibitemShut {NoStop}%
\bibitem [{\citenamefont {Charity}\ \emph
  {et~al.}(2021{\natexlab{a}})\citenamefont {Charity}, \citenamefont {Webb},
  \citenamefont {Elson}, \citenamefont {Hoff}, \citenamefont {Pruitt},
  \citenamefont {Sobotka}, \citenamefont {Brown}, \citenamefont {Cerizza},
  \citenamefont {Estee}, \citenamefont {Lynch}, \citenamefont {Manfredi},
  \citenamefont {Morfouace}, \citenamefont {Santamaria}, \citenamefont
  {Sweany}, \citenamefont {Tsang}, \citenamefont {Tsang}, \citenamefont
  {Zhang}, \citenamefont {Zhu}, \citenamefont {Kuvin}, \citenamefont {McNeel},
  \citenamefont {Smith}, \citenamefont {Wuosmaa},\ and\ \citenamefont
  {Chajecki}}]{Charity:2021}%
  \BibitemOpen
  \bibfield  {author} {\bibinfo {author} {\bibfnamefont {R.~J.}\ \bibnamefont
  {Charity}}, \bibinfo {author} {\bibfnamefont {T.~B.}\ \bibnamefont {Webb}},
  \bibinfo {author} {\bibfnamefont {J.~M.}\ \bibnamefont {Elson}}, \bibinfo
  {author} {\bibfnamefont {D.~E.~M.}\ \bibnamefont {Hoff}}, \bibinfo {author}
  {\bibfnamefont {C.~D.}\ \bibnamefont {Pruitt}}, \bibinfo {author}
  {\bibfnamefont {L.~G.}\ \bibnamefont {Sobotka}}, \bibinfo {author}
  {\bibfnamefont {K.~W.}\ \bibnamefont {Brown}}, \bibinfo {author}
  {\bibfnamefont {G.}~\bibnamefont {Cerizza}}, \bibinfo {author} {\bibfnamefont
  {J.}~\bibnamefont {Estee}}, \bibinfo {author} {\bibfnamefont {W.~G.}\
  \bibnamefont {Lynch}}, \bibinfo {author} {\bibfnamefont {J.}~\bibnamefont
  {Manfredi}}, \bibinfo {author} {\bibfnamefont {P.}~\bibnamefont {Morfouace}},
  \bibinfo {author} {\bibfnamefont {C.}~\bibnamefont {Santamaria}}, \bibinfo
  {author} {\bibfnamefont {S.}~\bibnamefont {Sweany}}, \bibinfo {author}
  {\bibfnamefont {C.~Y.}\ \bibnamefont {Tsang}}, \bibinfo {author}
  {\bibfnamefont {M.~B.}\ \bibnamefont {Tsang}}, \bibinfo {author}
  {\bibfnamefont {Y.}~\bibnamefont {Zhang}}, \bibinfo {author} {\bibfnamefont
  {K.}~\bibnamefont {Zhu}}, \bibinfo {author} {\bibfnamefont {S.~A.}\
  \bibnamefont {Kuvin}}, \bibinfo {author} {\bibfnamefont {D.}~\bibnamefont
  {McNeel}}, \bibinfo {author} {\bibfnamefont {J.}~\bibnamefont {Smith}},
  \bibinfo {author} {\bibfnamefont {A.~H.}\ \bibnamefont {Wuosmaa}},\ and\
  \bibinfo {author} {\bibfnamefont {Z.}~\bibnamefont {Chajecki}},\ }\bibfield
  {title} {\bibinfo {title} {Observation of the exotic isotope
  $^{13}\mathrm{F}$ located four neutrons beyond the proton drip line},\ }\href
  {https://doi.org/10.1103/PhysRevLett.126.132501} {\bibfield  {journal}
  {\bibinfo  {journal} {Phys. Rev. Lett.}\ }\textbf {\bibinfo {volume} {126}},\
  \bibinfo {pages} {132501} (\bibinfo {year} {2021}{\natexlab{a}})}\BibitemShut
  {NoStop}%
\bibitem [{\citenamefont {Brown}\ \emph {et~al.}(2017)\citenamefont {Brown},
  \citenamefont {Charity}, \citenamefont {Elson}, \citenamefont {Reviol},
  \citenamefont {Sobotka}, \citenamefont {Buhro}, \citenamefont {Chajecki},
  \citenamefont {Lynch}, \citenamefont {Manfredi}, \citenamefont {Shane},
  \citenamefont {Showalter}, \citenamefont {Tsang}, \citenamefont {Weisshaar},
  \citenamefont {Winkelbauer}, \citenamefont {Bedoor},\ and\ \citenamefont
  {Wuosmaa}}]{Brown:2017}%
  \BibitemOpen
  \bibfield  {author} {\bibinfo {author} {\bibfnamefont {K.~W.}\ \bibnamefont
  {Brown}}, \bibinfo {author} {\bibfnamefont {R.~J.}\ \bibnamefont {Charity}},
  \bibinfo {author} {\bibfnamefont {J.~M.}\ \bibnamefont {Elson}}, \bibinfo
  {author} {\bibfnamefont {W.}~\bibnamefont {Reviol}}, \bibinfo {author}
  {\bibfnamefont {L.~G.}\ \bibnamefont {Sobotka}}, \bibinfo {author}
  {\bibfnamefont {W.~W.}\ \bibnamefont {Buhro}}, \bibinfo {author}
  {\bibfnamefont {Z.}~\bibnamefont {Chajecki}}, \bibinfo {author}
  {\bibfnamefont {W.~G.}\ \bibnamefont {Lynch}}, \bibinfo {author}
  {\bibfnamefont {J.}~\bibnamefont {Manfredi}}, \bibinfo {author}
  {\bibfnamefont {R.}~\bibnamefont {Shane}}, \bibinfo {author} {\bibfnamefont
  {R.~H.}\ \bibnamefont {Showalter}}, \bibinfo {author} {\bibfnamefont {M.~B.}\
  \bibnamefont {Tsang}}, \bibinfo {author} {\bibfnamefont {D.}~\bibnamefont
  {Weisshaar}}, \bibinfo {author} {\bibfnamefont {J.~R.}\ \bibnamefont
  {Winkelbauer}}, \bibinfo {author} {\bibfnamefont {S.}~\bibnamefont
  {Bedoor}},\ and\ \bibinfo {author} {\bibfnamefont {A.~H.}\ \bibnamefont
  {Wuosmaa}},\ }\bibfield  {title} {\bibinfo {title} {Proton-decaying states in
  light nuclei and the first observation of $^{17}\mathrm{Na}$},\ }\href
  {https://doi.org/10.1103/PhysRevC.95.044326} {\bibfield  {journal} {\bibinfo
  {journal} {Phys. Rev. C}\ }\textbf {\bibinfo {volume} {95}},\ \bibinfo
  {pages} {044326} (\bibinfo {year} {2017})}\BibitemShut {NoStop}%
\bibitem [{\citenamefont {Kostyleva}\ \emph {et~al.}(2019)\citenamefont
  {Kostyleva}, \citenamefont {Mukha}, \citenamefont {Acosta}, \citenamefont
  {Casarejos}, \citenamefont {Chudoba}, \citenamefont {Ciemny}, \citenamefont
  {Dominik}, \citenamefont {Due\~nas}, \citenamefont {Dunin}, \citenamefont
  {Espino}, \citenamefont {Estrad\'e}, \citenamefont {Farinon}, \citenamefont
  {Fomichev}, \citenamefont {Geissel}, \citenamefont {Gorshkov}, \citenamefont
  {Grigorenko}, \citenamefont {Janas}, \citenamefont
  {Kami\ifmmode~\acute{n}\else \'{n}\fi{}ski}, \citenamefont {Kiselev},
  \citenamefont {Kn\"obel}, \citenamefont {Krupko}, \citenamefont {Kuich},
  \citenamefont {Litvinov}, \citenamefont {Marquinez-Dur\'an}, \citenamefont
  {Martel}, \citenamefont {Mazzocchi}, \citenamefont {Nociforo}, \citenamefont
  {Ord\'uz}, \citenamefont {Pf\"utzner}, \citenamefont {Pietri}, \citenamefont
  {Pomorski}, \citenamefont {Prochazka}, \citenamefont {Rymzhanova},
  \citenamefont {S\'anchez-Ben\'{\i}tez}, \citenamefont {Scheidenberger},
  \citenamefont {Simon}, \citenamefont {Sitar}, \citenamefont {Slepnev},
  \citenamefont {Stanoiu}, \citenamefont {Strmen}, \citenamefont {Szarka},
  \citenamefont {Takechi}, \citenamefont {Tanaka}, \citenamefont {Weick},
  \citenamefont {Winkler}, \citenamefont {Winfield}, \citenamefont {Xu},\ and\
  \citenamefont {Zhukov}}]{Kostyleva:2019}%
  \BibitemOpen
  \bibfield  {author} {\bibinfo {author} {\bibfnamefont {D.}~\bibnamefont
  {Kostyleva}}, \bibinfo {author} {\bibfnamefont {I.}~\bibnamefont {Mukha}},
  \bibinfo {author} {\bibfnamefont {L.}~\bibnamefont {Acosta}}, \bibinfo
  {author} {\bibfnamefont {E.}~\bibnamefont {Casarejos}}, \bibinfo {author}
  {\bibfnamefont {V.}~\bibnamefont {Chudoba}}, \bibinfo {author} {\bibfnamefont
  {A.~A.}\ \bibnamefont {Ciemny}}, \bibinfo {author} {\bibfnamefont
  {W.}~\bibnamefont {Dominik}}, \bibinfo {author} {\bibfnamefont {J.~A.}\
  \bibnamefont {Due\~nas}}, \bibinfo {author} {\bibfnamefont {V.}~\bibnamefont
  {Dunin}}, \bibinfo {author} {\bibfnamefont {J.~M.}\ \bibnamefont {Espino}},
  \bibinfo {author} {\bibfnamefont {A.}~\bibnamefont {Estrad\'e}}, \bibinfo
  {author} {\bibfnamefont {F.}~\bibnamefont {Farinon}}, \bibinfo {author}
  {\bibfnamefont {A.}~\bibnamefont {Fomichev}}, \bibinfo {author}
  {\bibfnamefont {H.}~\bibnamefont {Geissel}}, \bibinfo {author} {\bibfnamefont
  {A.}~\bibnamefont {Gorshkov}}, \bibinfo {author} {\bibfnamefont {L.~V.}\
  \bibnamefont {Grigorenko}}, \bibinfo {author} {\bibfnamefont
  {Z.}~\bibnamefont {Janas}}, \bibinfo {author} {\bibfnamefont
  {G.}~\bibnamefont {Kami\ifmmode~\acute{n}\else \'{n}\fi{}ski}}, \bibinfo
  {author} {\bibfnamefont {O.}~\bibnamefont {Kiselev}}, \bibinfo {author}
  {\bibfnamefont {R.}~\bibnamefont {Kn\"obel}}, \bibinfo {author}
  {\bibfnamefont {S.}~\bibnamefont {Krupko}}, \bibinfo {author} {\bibfnamefont
  {M.}~\bibnamefont {Kuich}}, \bibinfo {author} {\bibfnamefont {Y.~A.}\
  \bibnamefont {Litvinov}}, \bibinfo {author} {\bibfnamefont {G.}~\bibnamefont
  {Marquinez-Dur\'an}}, \bibinfo {author} {\bibfnamefont {I.}~\bibnamefont
  {Martel}}, \bibinfo {author} {\bibfnamefont {C.}~\bibnamefont {Mazzocchi}},
  \bibinfo {author} {\bibfnamefont {C.}~\bibnamefont {Nociforo}}, \bibinfo
  {author} {\bibfnamefont {A.~K.}\ \bibnamefont {Ord\'uz}}, \bibinfo {author}
  {\bibfnamefont {M.}~\bibnamefont {Pf\"utzner}}, \bibinfo {author}
  {\bibfnamefont {S.}~\bibnamefont {Pietri}}, \bibinfo {author} {\bibfnamefont
  {M.}~\bibnamefont {Pomorski}}, \bibinfo {author} {\bibfnamefont
  {A.}~\bibnamefont {Prochazka}}, \bibinfo {author} {\bibfnamefont
  {S.}~\bibnamefont {Rymzhanova}}, \bibinfo {author} {\bibfnamefont {A.~M.}\
  \bibnamefont {S\'anchez-Ben\'{\i}tez}}, \bibinfo {author} {\bibfnamefont
  {C.}~\bibnamefont {Scheidenberger}}, \bibinfo {author} {\bibfnamefont
  {H.}~\bibnamefont {Simon}}, \bibinfo {author} {\bibfnamefont
  {B.}~\bibnamefont {Sitar}}, \bibinfo {author} {\bibfnamefont
  {R.}~\bibnamefont {Slepnev}}, \bibinfo {author} {\bibfnamefont
  {M.}~\bibnamefont {Stanoiu}}, \bibinfo {author} {\bibfnamefont
  {P.}~\bibnamefont {Strmen}}, \bibinfo {author} {\bibfnamefont
  {I.}~\bibnamefont {Szarka}}, \bibinfo {author} {\bibfnamefont
  {M.}~\bibnamefont {Takechi}}, \bibinfo {author} {\bibfnamefont {Y.~K.}\
  \bibnamefont {Tanaka}}, \bibinfo {author} {\bibfnamefont {H.}~\bibnamefont
  {Weick}}, \bibinfo {author} {\bibfnamefont {M.}~\bibnamefont {Winkler}},
  \bibinfo {author} {\bibfnamefont {J.~S.}\ \bibnamefont {Winfield}}, \bibinfo
  {author} {\bibfnamefont {X.}~\bibnamefont {Xu}},\ and\ \bibinfo {author}
  {\bibfnamefont {M.~V.}\ \bibnamefont {Zhukov}},\ }\bibfield  {title}
  {\bibinfo {title} {Towards the limits of existence of nuclear structure:
  Observation and first spectroscopy of the isotope $^{31}\mathrm{K}$ by
  measuring its three-proton decay},\ }\href
  {https://doi.org/10.1103/PhysRevLett.123.092502} {\bibfield  {journal}
  {\bibinfo  {journal} {Phys. Rev. Lett.}\ }\textbf {\bibinfo {volume} {123}},\
  \bibinfo {pages} {092502} (\bibinfo {year} {2019})}\BibitemShut {NoStop}%
\bibitem [{\citenamefont {Charity}\ \emph {et~al.}(2010)\citenamefont
  {Charity}, \citenamefont {Elson}, \citenamefont {Manfredi}, \citenamefont
  {Shane}, \citenamefont {Sobotka}, \citenamefont {Chajecki}, \citenamefont
  {Coupland}, \citenamefont {Iwasaki}, \citenamefont {Kilburn}, \citenamefont
  {Lee}, \citenamefont {Lynch}, \citenamefont {Sanetullaev}, \citenamefont
  {Tsang}, \citenamefont {Winkelbauer}, \citenamefont {Youngs}, \citenamefont
  {Marley}, \citenamefont {Shetty}, \citenamefont {Wuosmaa}, \citenamefont
  {Ghosh},\ and\ \citenamefont {Howard}}]{Charity:2010}%
  \BibitemOpen
  \bibfield  {author} {\bibinfo {author} {\bibfnamefont {R.~J.}\ \bibnamefont
  {Charity}}, \bibinfo {author} {\bibfnamefont {J.~M.}\ \bibnamefont {Elson}},
  \bibinfo {author} {\bibfnamefont {J.}~\bibnamefont {Manfredi}}, \bibinfo
  {author} {\bibfnamefont {R.}~\bibnamefont {Shane}}, \bibinfo {author}
  {\bibfnamefont {L.~G.}\ \bibnamefont {Sobotka}}, \bibinfo {author}
  {\bibfnamefont {Z.}~\bibnamefont {Chajecki}}, \bibinfo {author}
  {\bibfnamefont {D.}~\bibnamefont {Coupland}}, \bibinfo {author}
  {\bibfnamefont {H.}~\bibnamefont {Iwasaki}}, \bibinfo {author} {\bibfnamefont
  {M.}~\bibnamefont {Kilburn}}, \bibinfo {author} {\bibfnamefont
  {J.}~\bibnamefont {Lee}}, \bibinfo {author} {\bibfnamefont {W.~G.}\
  \bibnamefont {Lynch}}, \bibinfo {author} {\bibfnamefont {A.}~\bibnamefont
  {Sanetullaev}}, \bibinfo {author} {\bibfnamefont {M.~B.}\ \bibnamefont
  {Tsang}}, \bibinfo {author} {\bibfnamefont {J.}~\bibnamefont {Winkelbauer}},
  \bibinfo {author} {\bibfnamefont {M.}~\bibnamefont {Youngs}}, \bibinfo
  {author} {\bibfnamefont {S.~T.}\ \bibnamefont {Marley}}, \bibinfo {author}
  {\bibfnamefont {D.~V.}\ \bibnamefont {Shetty}}, \bibinfo {author}
  {\bibfnamefont {A.~H.}\ \bibnamefont {Wuosmaa}}, \bibinfo {author}
  {\bibfnamefont {T.~K.}\ \bibnamefont {Ghosh}},\ and\ \bibinfo {author}
  {\bibfnamefont {M.~E.}\ \bibnamefont {Howard}},\ }\bibfield  {title}
  {\bibinfo {title} {$2p$-$2p$ decay of ${}^{8}\mathbf{C}$ and isospin-allowed
  $2p$ decay of the isobaric-analog state in ${}^{8}\mathbf{B}$},\ }\href
  {https://doi.org/10.1103/PhysRevC.82.041304} {\bibfield  {journal} {\bibinfo
  {journal} {Phys. Rev. C}\ }\textbf {\bibinfo {volume} {82}},\ \bibinfo
  {pages} {041304(R)} (\bibinfo {year} {2010})}\BibitemShut {NoStop}%
\bibitem [{\citenamefont {Jin}\ \emph {et~al.}(2021)\citenamefont {Jin},
  \citenamefont {Niu}, \citenamefont {Brown}, \citenamefont {Li}, \citenamefont
  {Hua}, \citenamefont {Anthony}, \citenamefont {Barney}, \citenamefont
  {Charity}, \citenamefont {Crosby}, \citenamefont {Dell'Aquila}, \citenamefont
  {Elson}, \citenamefont {Estee}, \citenamefont {Ghazali}, \citenamefont
  {Jhang}, \citenamefont {Li}, \citenamefont {Lynch}, \citenamefont {Michel},
  \citenamefont {Sobotka}, \citenamefont {Sweany}, \citenamefont {Teh},
  \citenamefont {Thomas}, \citenamefont {Tsang}, \citenamefont {Tsang},
  \citenamefont {Wang}, \citenamefont {Wu}, \citenamefont {Yuan},\ and\
  \citenamefont {Zhu}}]{Jin:2021}%
  \BibitemOpen
  \bibfield  {author} {\bibinfo {author} {\bibfnamefont {Y.}~\bibnamefont
  {Jin}}, \bibinfo {author} {\bibfnamefont {C.~Y.}\ \bibnamefont {Niu}},
  \bibinfo {author} {\bibfnamefont {K.~W.}\ \bibnamefont {Brown}}, \bibinfo
  {author} {\bibfnamefont {Z.~H.}\ \bibnamefont {Li}}, \bibinfo {author}
  {\bibfnamefont {H.}~\bibnamefont {Hua}}, \bibinfo {author} {\bibfnamefont
  {A.~K.}\ \bibnamefont {Anthony}}, \bibinfo {author} {\bibfnamefont
  {J.}~\bibnamefont {Barney}}, \bibinfo {author} {\bibfnamefont {R.~J.}\
  \bibnamefont {Charity}}, \bibinfo {author} {\bibfnamefont {J.}~\bibnamefont
  {Crosby}}, \bibinfo {author} {\bibfnamefont {D.}~\bibnamefont {Dell'Aquila}},
  \bibinfo {author} {\bibfnamefont {J.~M.}\ \bibnamefont {Elson}}, \bibinfo
  {author} {\bibfnamefont {J.}~\bibnamefont {Estee}}, \bibinfo {author}
  {\bibfnamefont {M.}~\bibnamefont {Ghazali}}, \bibinfo {author} {\bibfnamefont
  {G.}~\bibnamefont {Jhang}}, \bibinfo {author} {\bibfnamefont {J.~G.}\
  \bibnamefont {Li}}, \bibinfo {author} {\bibfnamefont {W.~G.}\ \bibnamefont
  {Lynch}}, \bibinfo {author} {\bibfnamefont {N.}~\bibnamefont {Michel}},
  \bibinfo {author} {\bibfnamefont {L.~G.}\ \bibnamefont {Sobotka}}, \bibinfo
  {author} {\bibfnamefont {S.}~\bibnamefont {Sweany}}, \bibinfo {author}
  {\bibfnamefont {F.~C.~E.}\ \bibnamefont {Teh}}, \bibinfo {author}
  {\bibfnamefont {A.}~\bibnamefont {Thomas}}, \bibinfo {author} {\bibfnamefont
  {C.~Y.}\ \bibnamefont {Tsang}}, \bibinfo {author} {\bibfnamefont {M.~B.}\
  \bibnamefont {Tsang}}, \bibinfo {author} {\bibfnamefont {S.~M.}\ \bibnamefont
  {Wang}}, \bibinfo {author} {\bibfnamefont {H.~Y.}\ \bibnamefont {Wu}},
  \bibinfo {author} {\bibfnamefont {C.~X.}\ \bibnamefont {Yuan}},\ and\
  \bibinfo {author} {\bibfnamefont {K.}~\bibnamefont {Zhu}},\ }\bibfield
  {title} {\bibinfo {title} {First observation of the four-proton unbound
  nucleus $^{18}\mathrm{Mg}$},\ }\href
  {https://doi.org/10.1103/PhysRevLett.127.262502} {\bibfield  {journal}
  {\bibinfo  {journal} {Phys. Rev. Lett.}\ }\textbf {\bibinfo {volume} {127}},\
  \bibinfo {pages} {262502} (\bibinfo {year} {2021})}\BibitemShut {NoStop}%
\bibitem [{\citenamefont {Thoennessen}(2004)}]{Thoennessen2004}%
  \BibitemOpen
  \bibfield  {author} {\bibinfo {author} {\bibfnamefont {M.}~\bibnamefont
  {Thoennessen}},\ }\bibfield  {title} {\bibinfo {title} {Reaching the limits
  of nuclear stability},\ }\href {https://doi.org/10.1088/0034-4885/67/7/r04}
  {\bibfield  {journal} {\bibinfo  {journal} {Rep. Prog. Phys.}\ }\textbf
  {\bibinfo {volume} {67}},\ \bibinfo {pages} {1187} (\bibinfo {year}
  {2004})}\BibitemShut {NoStop}%
\bibitem [{\citenamefont {Fossez}\ \emph {et~al.}(2016)\citenamefont {Fossez},
  \citenamefont {Nazarewicz}, \citenamefont {Jaganathen}, \citenamefont
  {Michel},\ and\ \citenamefont {P\l{}oszajczak}}]{Fossez:2016}%
  \BibitemOpen
  \bibfield  {author} {\bibinfo {author} {\bibfnamefont {K.}~\bibnamefont
  {Fossez}}, \bibinfo {author} {\bibfnamefont {W.}~\bibnamefont {Nazarewicz}},
  \bibinfo {author} {\bibfnamefont {Y.}~\bibnamefont {Jaganathen}}, \bibinfo
  {author} {\bibfnamefont {N.}~\bibnamefont {Michel}},\ and\ \bibinfo {author}
  {\bibfnamefont {M.}~\bibnamefont {P\l{}oszajczak}},\ }\bibfield  {title}
  {\bibinfo {title} {Nuclear rotation in the continuum},\ }\href
  {https://doi.org/10.1103/PhysRevC.93.011305} {\bibfield  {journal} {\bibinfo
  {journal} {Phys. Rev. C}\ }\textbf {\bibinfo {volume} {93}},\ \bibinfo
  {pages} {011305(R)} (\bibinfo {year} {2016})}\BibitemShut {NoStop}%
\bibitem [{\citenamefont {Ohanian}\ and\ \citenamefont
  {Ginsburg}(1974)}]{Ohanian1974}%
  \BibitemOpen
  \bibfield  {author} {\bibinfo {author} {\bibfnamefont {H.~C.}\ \bibnamefont
  {Ohanian}}\ and\ \bibinfo {author} {\bibfnamefont {C.~G.}\ \bibnamefont
  {Ginsburg}},\ }\bibfield  {title} {\bibinfo {title} {Antibound 'states' and
  resonances},\ }\href {https://doi.org/10.1119/1.1987678} {\bibfield
  {journal} {\bibinfo  {journal} {Am. J. Phys.}\ }\textbf {\bibinfo {volume}
  {42}},\ \bibinfo {pages} {310} (\bibinfo {year} {1974})}\BibitemShut
  {NoStop}%
\bibitem [{\citenamefont {Babenko}\ and\ \citenamefont
  {Petrov}(2013)}]{Babenko:2013}%
  \BibitemOpen
  \bibfield  {author} {\bibinfo {author} {\bibfnamefont {V.~A.}\ \bibnamefont
  {Babenko}}\ and\ \bibinfo {author} {\bibfnamefont {N.~M.}\ \bibnamefont
  {Petrov}},\ }\bibfield  {title} {\bibinfo {title} {Low-energy parameters of
  neutron-neutron interaction in the effective-range approximation},\ }\href
  {https://doi.org/10.1134/S1063778813060033} {\bibfield  {journal} {\bibinfo
  {journal} {Phys. Atom. Nuclei}\ }\textbf {\bibinfo {volume} {76}},\ \bibinfo
  {pages} {684} (\bibinfo {year} {2013})}\BibitemShut {NoStop}%
\bibitem [{\citenamefont {Kok}(1980)}]{Kok:1980}%
  \BibitemOpen
  \bibfield  {author} {\bibinfo {author} {\bibfnamefont {L.~P.}\ \bibnamefont
  {Kok}},\ }\bibfield  {title} {\bibinfo {title} {Accurate determination of the
  ground-state level of the $^{2}\mathrm{He}$ nucleus},\ }\href
  {https://doi.org/10.1103/PhysRevLett.45.427} {\bibfield  {journal} {\bibinfo
  {journal} {Phys. Rev. Lett.}\ }\textbf {\bibinfo {volume} {45}},\ \bibinfo
  {pages} {427} (\bibinfo {year} {1980})}\BibitemShut {NoStop}%
\bibitem [{\citenamefont {Mukhamedzhanov}\ \emph {et~al.}(2010)\citenamefont
  {Mukhamedzhanov}, \citenamefont {Irgaziev}, \citenamefont {Goldberg},
  \citenamefont {Orlov},\ and\ \citenamefont {Qazi}}]{Mukhamedzhanov2010}%
  \BibitemOpen
  \bibfield  {author} {\bibinfo {author} {\bibfnamefont {A.~M.}\ \bibnamefont
  {Mukhamedzhanov}}, \bibinfo {author} {\bibfnamefont {B.~F.}\ \bibnamefont
  {Irgaziev}}, \bibinfo {author} {\bibfnamefont {V.~Z.}\ \bibnamefont
  {Goldberg}}, \bibinfo {author} {\bibfnamefont {Y.~V.}\ \bibnamefont
  {Orlov}},\ and\ \bibinfo {author} {\bibfnamefont {I.}~\bibnamefont {Qazi}},\
  }\bibfield  {title} {\bibinfo {title} {Bound, virtual, and resonance
  {$S$-matrix} poles from the {Schr\"odinger} equation},\ }\href
  {https://doi.org/10.1103/PhysRevC.81.054314} {\bibfield  {journal} {\bibinfo
  {journal} {Phys. Rev. C}\ }\textbf {\bibinfo {volume} {81}},\ \bibinfo
  {pages} {054314} (\bibinfo {year} {2010})}\BibitemShut {NoStop}%
\bibitem [{\citenamefont {Kisamori}\ \emph {et~al.}(2016)\citenamefont
  {Kisamori}, \citenamefont {Shimoura}, \citenamefont {Miya}, \citenamefont
  {Michimasa}, \citenamefont {Ota}, \citenamefont {Assie}, \citenamefont
  {Baba}, \citenamefont {Baba}, \citenamefont {Beaumel}, \citenamefont
  {Dozono}, \citenamefont {Fujii}, \citenamefont {Fukuda}, \citenamefont {Go},
  \citenamefont {Hammache}, \citenamefont {Ideguchi}, \citenamefont {Inabe},
  \citenamefont {Itoh}, \citenamefont {Kameda}, \citenamefont {Kawase},
  \citenamefont {Kawabata}, \citenamefont {Kobayashi}, \citenamefont {Kondo},
  \citenamefont {Kubo}, \citenamefont {Kubota}, \citenamefont
  {Kurata-Nishimura}, \citenamefont {Lee}, \citenamefont {Maeda}, \citenamefont
  {Matsubara}, \citenamefont {Miki}, \citenamefont {Nishi}, \citenamefont
  {Noji}, \citenamefont {Sakaguchi}, \citenamefont {Sakai}, \citenamefont
  {Sasamoto}, \citenamefont {Sasano}, \citenamefont {Sato}, \citenamefont
  {Shimizu}, \citenamefont {Stolz}, \citenamefont {Suzuki}, \citenamefont
  {Takaki}, \citenamefont {Takeda}, \citenamefont {Takeuchi}, \citenamefont
  {Tamii}, \citenamefont {Tang}, \citenamefont {Tokieda}, \citenamefont
  {Tsumura}, \citenamefont {Uesaka}, \citenamefont {Yako}, \citenamefont
  {Yanagisawa}, \citenamefont {Yokoyama},\ and\ \citenamefont
  {Yoshida}}]{Kisamori:2016}%
  \BibitemOpen
  \bibfield  {author} {\bibinfo {author} {\bibfnamefont {K.}~\bibnamefont
  {Kisamori}}, \bibinfo {author} {\bibfnamefont {S.}~\bibnamefont {Shimoura}},
  \bibinfo {author} {\bibfnamefont {H.}~\bibnamefont {Miya}}, \bibinfo {author}
  {\bibfnamefont {S.}~\bibnamefont {Michimasa}}, \bibinfo {author}
  {\bibfnamefont {S.}~\bibnamefont {Ota}}, \bibinfo {author} {\bibfnamefont
  {M.}~\bibnamefont {Assie}}, \bibinfo {author} {\bibfnamefont
  {H.}~\bibnamefont {Baba}}, \bibinfo {author} {\bibfnamefont {T.}~\bibnamefont
  {Baba}}, \bibinfo {author} {\bibfnamefont {D.}~\bibnamefont {Beaumel}},
  \bibinfo {author} {\bibfnamefont {M.}~\bibnamefont {Dozono}}, \bibinfo
  {author} {\bibfnamefont {T.}~\bibnamefont {Fujii}}, \bibinfo {author}
  {\bibfnamefont {N.}~\bibnamefont {Fukuda}}, \bibinfo {author} {\bibfnamefont
  {S.}~\bibnamefont {Go}}, \bibinfo {author} {\bibfnamefont {F.}~\bibnamefont
  {Hammache}}, \bibinfo {author} {\bibfnamefont {E.}~\bibnamefont {Ideguchi}},
  \bibinfo {author} {\bibfnamefont {N.}~\bibnamefont {Inabe}}, \bibinfo
  {author} {\bibfnamefont {M.}~\bibnamefont {Itoh}}, \bibinfo {author}
  {\bibfnamefont {D.}~\bibnamefont {Kameda}}, \bibinfo {author} {\bibfnamefont
  {S.}~\bibnamefont {Kawase}}, \bibinfo {author} {\bibfnamefont
  {T.}~\bibnamefont {Kawabata}}, \bibinfo {author} {\bibfnamefont
  {M.}~\bibnamefont {Kobayashi}}, \bibinfo {author} {\bibfnamefont
  {Y.}~\bibnamefont {Kondo}}, \bibinfo {author} {\bibfnamefont
  {T.}~\bibnamefont {Kubo}}, \bibinfo {author} {\bibfnamefont {Y.}~\bibnamefont
  {Kubota}}, \bibinfo {author} {\bibfnamefont {M.}~\bibnamefont
  {Kurata-Nishimura}}, \bibinfo {author} {\bibfnamefont {C.~S.}\ \bibnamefont
  {Lee}}, \bibinfo {author} {\bibfnamefont {Y.}~\bibnamefont {Maeda}}, \bibinfo
  {author} {\bibfnamefont {H.}~\bibnamefont {Matsubara}}, \bibinfo {author}
  {\bibfnamefont {K.}~\bibnamefont {Miki}}, \bibinfo {author} {\bibfnamefont
  {T.}~\bibnamefont {Nishi}}, \bibinfo {author} {\bibfnamefont
  {S.}~\bibnamefont {Noji}}, \bibinfo {author} {\bibfnamefont {S.}~\bibnamefont
  {Sakaguchi}}, \bibinfo {author} {\bibfnamefont {H.}~\bibnamefont {Sakai}},
  \bibinfo {author} {\bibfnamefont {Y.}~\bibnamefont {Sasamoto}}, \bibinfo
  {author} {\bibfnamefont {M.}~\bibnamefont {Sasano}}, \bibinfo {author}
  {\bibfnamefont {H.}~\bibnamefont {Sato}}, \bibinfo {author} {\bibfnamefont
  {Y.}~\bibnamefont {Shimizu}}, \bibinfo {author} {\bibfnamefont
  {A.}~\bibnamefont {Stolz}}, \bibinfo {author} {\bibfnamefont
  {H.}~\bibnamefont {Suzuki}}, \bibinfo {author} {\bibfnamefont
  {M.}~\bibnamefont {Takaki}}, \bibinfo {author} {\bibfnamefont
  {H.}~\bibnamefont {Takeda}}, \bibinfo {author} {\bibfnamefont
  {S.}~\bibnamefont {Takeuchi}}, \bibinfo {author} {\bibfnamefont
  {A.}~\bibnamefont {Tamii}}, \bibinfo {author} {\bibfnamefont
  {L.}~\bibnamefont {Tang}}, \bibinfo {author} {\bibfnamefont {H.}~\bibnamefont
  {Tokieda}}, \bibinfo {author} {\bibfnamefont {M.}~\bibnamefont {Tsumura}},
  \bibinfo {author} {\bibfnamefont {T.}~\bibnamefont {Uesaka}}, \bibinfo
  {author} {\bibfnamefont {K.}~\bibnamefont {Yako}}, \bibinfo {author}
  {\bibfnamefont {Y.}~\bibnamefont {Yanagisawa}}, \bibinfo {author}
  {\bibfnamefont {R.}~\bibnamefont {Yokoyama}},\ and\ \bibinfo {author}
  {\bibfnamefont {K.}~\bibnamefont {Yoshida}},\ }\bibfield  {title} {\bibinfo
  {title} {Candidate resonant tetraneutron state populated by the
  $^{4}\mathrm{He}(^{8}\mathrm{He},^{8}\mathrm{Be})$ reaction},\ }\href
  {https://doi.org/10.1103/PhysRevLett.116.052501} {\bibfield  {journal}
  {\bibinfo  {journal} {Phys. Rev. Lett.}\ }\textbf {\bibinfo {volume} {116}},\
  \bibinfo {pages} {052501} (\bibinfo {year} {2016})}\BibitemShut {NoStop}%
\bibitem [{\citenamefont {Duer}\ \emph {et~al.}(2022)\citenamefont {Duer} \emph
  {et~al.}}]{Duer:2022}%
  \BibitemOpen
  \bibfield  {author} {\bibinfo {author} {\bibfnamefont {M.}~\bibnamefont
  {Duer}} \emph {et~al.},\ }\bibfield  {title} {\bibinfo {title} {Observation
  of a correlated free four-neutron system},\ }\href
  {https://doi.org/10.1038/s41586-022-04827-6} {\bibfield  {journal} {\bibinfo
  {journal} {Nature}\ }\textbf {\bibinfo {volume} {606}},\ \bibinfo {pages}
  {678} (\bibinfo {year} {2022})}\BibitemShut {NoStop}%
\bibitem [{\citenamefont {Deltuva}(2018)}]{Deltuva:2018}%
  \BibitemOpen
  \bibfield  {author} {\bibinfo {author} {\bibfnamefont {A.}~\bibnamefont
  {Deltuva}},\ }\bibfield  {title} {\bibinfo {title} {Tetraneutron: Rigorous
  continuum calculation},\ }\href
  {https://doi.org/10.1016/j.physletb.2018.05.041} {\bibfield  {journal}
  {\bibinfo  {journal} {Phys. Lett. B}\ }\textbf {\bibinfo {volume} {782}},\
  \bibinfo {pages} {238} (\bibinfo {year} {2018})}\BibitemShut {NoStop}%
\bibitem [{\citenamefont {Higgins}\ \emph {et~al.}(2020)\citenamefont
  {Higgins}, \citenamefont {Greene}, \citenamefont {Kievsky},\ and\
  \citenamefont {Viviani}}]{Higgins:2020}%
  \BibitemOpen
  \bibfield  {author} {\bibinfo {author} {\bibfnamefont {M.~D.}\ \bibnamefont
  {Higgins}}, \bibinfo {author} {\bibfnamefont {C.~H.}\ \bibnamefont {Greene}},
  \bibinfo {author} {\bibfnamefont {A.}~\bibnamefont {Kievsky}},\ and\ \bibinfo
  {author} {\bibfnamefont {M.}~\bibnamefont {Viviani}},\ }\bibfield  {title}
  {\bibinfo {title} {Nonresonant density of states enhancement at low energies
  for three or four neutrons},\ }\href
  {https://doi.org/10.1103/PhysRevLett.125.052501} {\bibfield  {journal}
  {\bibinfo  {journal} {Phys. Rev. Lett.}\ }\textbf {\bibinfo {volume} {125}},\
  \bibinfo {pages} {052501} (\bibinfo {year} {2020})}\BibitemShut {NoStop}%
\bibitem [{\citenamefont {L\'epine-Szily}\ \emph {et~al.}(2002)\citenamefont
  {L\'epine-Szily}, \citenamefont {Oliveira}, \citenamefont {Vanin},
  \citenamefont {Ostrowski}, \citenamefont {Lichtenth\"aler}, \citenamefont
  {Di~Pietro}, \citenamefont {Guimar\~aes}, \citenamefont {Laird},
  \citenamefont {Maunoury}, \citenamefont {Lima}, \citenamefont
  {de~Oliveira~Santos}, \citenamefont {Roussel-Chomaz}, \citenamefont
  {Savajols}, \citenamefont {Trinder}, \citenamefont {Villari},\ and\
  \citenamefont {de~Vismes}}]{Lepine-Szily:2002}%
  \BibitemOpen
  \bibfield  {author} {\bibinfo {author} {\bibfnamefont {A.}~\bibnamefont
  {L\'epine-Szily}}, \bibinfo {author} {\bibfnamefont {J.~M.}\ \bibnamefont
  {Oliveira}}, \bibinfo {author} {\bibfnamefont {V.~R.}\ \bibnamefont {Vanin}},
  \bibinfo {author} {\bibfnamefont {A.~N.}\ \bibnamefont {Ostrowski}}, \bibinfo
  {author} {\bibfnamefont {R.}~\bibnamefont {Lichtenth\"aler}}, \bibinfo
  {author} {\bibfnamefont {A.}~\bibnamefont {Di~Pietro}}, \bibinfo {author}
  {\bibfnamefont {V.}~\bibnamefont {Guimar\~aes}}, \bibinfo {author}
  {\bibfnamefont {A.~M.}\ \bibnamefont {Laird}}, \bibinfo {author}
  {\bibfnamefont {L.}~\bibnamefont {Maunoury}}, \bibinfo {author}
  {\bibfnamefont {G.~F.}\ \bibnamefont {Lima}}, \bibinfo {author}
  {\bibfnamefont {F.}~\bibnamefont {de~Oliveira~Santos}}, \bibinfo {author}
  {\bibfnamefont {P.}~\bibnamefont {Roussel-Chomaz}}, \bibinfo {author}
  {\bibfnamefont {H.}~\bibnamefont {Savajols}}, \bibinfo {author}
  {\bibfnamefont {W.}~\bibnamefont {Trinder}}, \bibinfo {author} {\bibfnamefont
  {A.~C.~C.}\ \bibnamefont {Villari}},\ and\ \bibinfo {author} {\bibfnamefont
  {A.}~\bibnamefont {de~Vismes}},\ }\bibfield  {title} {\bibinfo {title}
  {Observation of the particle-unstable nucleus ${}^{10}\mathrm{N}$},\ }\href
  {https://doi.org/10.1103/PhysRevC.65.054318} {\bibfield  {journal} {\bibinfo
  {journal} {Phys. Rev. C}\ }\textbf {\bibinfo {volume} {65}},\ \bibinfo
  {pages} {054318} (\bibinfo {year} {2002})}\BibitemShut {NoStop}%
\bibitem [{\citenamefont {Hooker}\ \emph {et~al.}(2017)\citenamefont {Hooker},
  \citenamefont {Rogachev}, \citenamefont {Goldberg}, \citenamefont {Koshchiy},
  \citenamefont {Roeder}, \citenamefont {Jayatissa}, \citenamefont {Hunt},
  \citenamefont {Magana}, \citenamefont {Upadhyayula}, \citenamefont
  {Uberseder},\ and\ \citenamefont {Saastamoinen}}]{Hooker:2017}%
  \BibitemOpen
  \bibfield  {author} {\bibinfo {author} {\bibfnamefont {J.}~\bibnamefont
  {Hooker}}, \bibinfo {author} {\bibfnamefont {G.}~\bibnamefont {Rogachev}},
  \bibinfo {author} {\bibfnamefont {V.}~\bibnamefont {Goldberg}}, \bibinfo
  {author} {\bibfnamefont {E.}~\bibnamefont {Koshchiy}}, \bibinfo {author}
  {\bibfnamefont {B.}~\bibnamefont {Roeder}}, \bibinfo {author} {\bibfnamefont
  {H.}~\bibnamefont {Jayatissa}}, \bibinfo {author} {\bibfnamefont
  {C.}~\bibnamefont {Hunt}}, \bibinfo {author} {\bibfnamefont {C.}~\bibnamefont
  {Magana}}, \bibinfo {author} {\bibfnamefont {S.}~\bibnamefont {Upadhyayula}},
  \bibinfo {author} {\bibfnamefont {E.}~\bibnamefont {Uberseder}},\ and\
  \bibinfo {author} {\bibfnamefont {A.}~\bibnamefont {Saastamoinen}},\
  }\bibfield  {title} {\bibinfo {title} {Structure of {$^{10}$N} in {$^{9}$C}+p
  resonance scattering},\ }\href
  {https://doi.org/https://doi.org/10.1016/j.physletb.2017.03.025} {\bibfield
  {journal} {\bibinfo  {journal} {Phys. Lett. B}\ }\textbf {\bibinfo {volume}
  {769}},\ \bibinfo {pages} {62 } (\bibinfo {year} {2017})}\BibitemShut
  {NoStop}%
\bibitem [{\citenamefont {Charity}\ \emph
  {et~al.}(2021{\natexlab{b}})\citenamefont {Charity}, \citenamefont {Webb},
  \citenamefont {Sobotka},\ and\ \citenamefont {Brown}}]{Charity:2021b}%
  \BibitemOpen
  \bibfield  {author} {\bibinfo {author} {\bibfnamefont {R.~J.}\ \bibnamefont
  {Charity}}, \bibinfo {author} {\bibfnamefont {T.~B.}\ \bibnamefont {Webb}},
  \bibinfo {author} {\bibfnamefont {L.~G.}\ \bibnamefont {Sobotka}},\ and\
  \bibinfo {author} {\bibfnamefont {K.~W.}\ \bibnamefont {Brown}},\ }\bibfield
  {title} {\bibinfo {title} {Spectroscopy of $^{10}\mathrm{N}$ with the
  invariant-mass method},\ }\href {https://doi.org/10.1103/PhysRevC.104.054307}
  {\bibfield  {journal} {\bibinfo  {journal} {Phys. Rev. C}\ }\textbf {\bibinfo
  {volume} {104}},\ \bibinfo {pages} {054307} (\bibinfo {year}
  {2021}{\natexlab{b}})}\BibitemShut {NoStop}%
\bibitem [{\citenamefont {Aumann}\ \emph {et~al.}(2000)\citenamefont {Aumann},
  \citenamefont {Navin}, \citenamefont {Balamuth}, \citenamefont {Bazin},
  \citenamefont {Blank}, \citenamefont {Brown}, \citenamefont {Bush},
  \citenamefont {Caggiano}, \citenamefont {Davids}, \citenamefont {Glasmacher},
  \citenamefont {Guimar\~aes}, \citenamefont {Hansen}, \citenamefont
  {Ibbotson}, \citenamefont {Karnes}, \citenamefont {Kolata}, \citenamefont
  {Maddalena}, \citenamefont {Pritychenko}, \citenamefont {Scheit},
  \citenamefont {Sherrill},\ and\ \citenamefont {Tostevin}}]{Aumann:2000}%
  \BibitemOpen
  \bibfield  {author} {\bibinfo {author} {\bibfnamefont {T.}~\bibnamefont
  {Aumann}}, \bibinfo {author} {\bibfnamefont {A.}~\bibnamefont {Navin}},
  \bibinfo {author} {\bibfnamefont {D.~P.}\ \bibnamefont {Balamuth}}, \bibinfo
  {author} {\bibfnamefont {D.}~\bibnamefont {Bazin}}, \bibinfo {author}
  {\bibfnamefont {B.}~\bibnamefont {Blank}}, \bibinfo {author} {\bibfnamefont
  {B.~A.}\ \bibnamefont {Brown}}, \bibinfo {author} {\bibfnamefont {J.~E.}\
  \bibnamefont {Bush}}, \bibinfo {author} {\bibfnamefont {J.~A.}\ \bibnamefont
  {Caggiano}}, \bibinfo {author} {\bibfnamefont {B.}~\bibnamefont {Davids}},
  \bibinfo {author} {\bibfnamefont {T.}~\bibnamefont {Glasmacher}}, \bibinfo
  {author} {\bibfnamefont {V.}~\bibnamefont {Guimar\~aes}}, \bibinfo {author}
  {\bibfnamefont {P.~G.}\ \bibnamefont {Hansen}}, \bibinfo {author}
  {\bibfnamefont {R.~W.}\ \bibnamefont {Ibbotson}}, \bibinfo {author}
  {\bibfnamefont {D.}~\bibnamefont {Karnes}}, \bibinfo {author} {\bibfnamefont
  {J.~J.}\ \bibnamefont {Kolata}}, \bibinfo {author} {\bibfnamefont
  {V.}~\bibnamefont {Maddalena}}, \bibinfo {author} {\bibfnamefont
  {B.}~\bibnamefont {Pritychenko}}, \bibinfo {author} {\bibfnamefont
  {H.}~\bibnamefont {Scheit}}, \bibinfo {author} {\bibfnamefont {B.~M.}\
  \bibnamefont {Sherrill}},\ and\ \bibinfo {author} {\bibfnamefont {J.~A.}\
  \bibnamefont {Tostevin}},\ }\bibfield  {title} {\bibinfo {title} {One-neutron
  knockout from individual single-particle states of {$^{11}$Be}},\ }\href
  {https://doi.org/10.1103/PhysRevLett.84.35} {\bibfield  {journal} {\bibinfo
  {journal} {Phys. Rev. Lett.}\ }\textbf {\bibinfo {volume} {84}},\ \bibinfo
  {pages} {35} (\bibinfo {year} {2000})}\BibitemShut {NoStop}%
\bibitem [{\citenamefont {Al~Kalanee}\ \emph {et~al.}(2013)\citenamefont
  {Al~Kalanee}, \citenamefont {Gibelin}, \citenamefont {Roussel-Chomaz},
  \citenamefont {Keeley}, \citenamefont {Beaumel}, \citenamefont {Blumenfeld},
  \citenamefont {Fern\'andez-Dom\'{\i}nguez}, \citenamefont {Force},
  \citenamefont {Gaudefroy}, \citenamefont {Gillibert}, \citenamefont
  {Guillot}, \citenamefont {Iwasaki}, \citenamefont {Krupko}, \citenamefont
  {Lapoux}, \citenamefont {Mittig}, \citenamefont {Mougeot}, \citenamefont
  {Nalpas}, \citenamefont {Pollacco}, \citenamefont {Rusek}, \citenamefont
  {Roger}, \citenamefont {Savajols}, \citenamefont {de~S\'er\'eville},
  \citenamefont {Sidorchuk}, \citenamefont {Suzuki}, \citenamefont {Strojek},\
  and\ \citenamefont {Orr}}]{Kalanee:2013}%
  \BibitemOpen
  \bibfield  {author} {\bibinfo {author} {\bibfnamefont {T.}~\bibnamefont
  {Al~Kalanee}}, \bibinfo {author} {\bibfnamefont {J.}~\bibnamefont {Gibelin}},
  \bibinfo {author} {\bibfnamefont {P.}~\bibnamefont {Roussel-Chomaz}},
  \bibinfo {author} {\bibfnamefont {N.}~\bibnamefont {Keeley}}, \bibinfo
  {author} {\bibfnamefont {D.}~\bibnamefont {Beaumel}}, \bibinfo {author}
  {\bibfnamefont {Y.}~\bibnamefont {Blumenfeld}}, \bibinfo {author}
  {\bibfnamefont {B.}~\bibnamefont {Fern\'andez-Dom\'{\i}nguez}}, \bibinfo
  {author} {\bibfnamefont {C.}~\bibnamefont {Force}}, \bibinfo {author}
  {\bibfnamefont {L.}~\bibnamefont {Gaudefroy}}, \bibinfo {author}
  {\bibfnamefont {A.}~\bibnamefont {Gillibert}}, \bibinfo {author}
  {\bibfnamefont {J.}~\bibnamefont {Guillot}}, \bibinfo {author} {\bibfnamefont
  {H.}~\bibnamefont {Iwasaki}}, \bibinfo {author} {\bibfnamefont
  {S.}~\bibnamefont {Krupko}}, \bibinfo {author} {\bibfnamefont
  {V.}~\bibnamefont {Lapoux}}, \bibinfo {author} {\bibfnamefont
  {W.}~\bibnamefont {Mittig}}, \bibinfo {author} {\bibfnamefont
  {X.}~\bibnamefont {Mougeot}}, \bibinfo {author} {\bibfnamefont
  {L.}~\bibnamefont {Nalpas}}, \bibinfo {author} {\bibfnamefont
  {E.}~\bibnamefont {Pollacco}}, \bibinfo {author} {\bibfnamefont
  {K.}~\bibnamefont {Rusek}}, \bibinfo {author} {\bibfnamefont
  {T.}~\bibnamefont {Roger}}, \bibinfo {author} {\bibfnamefont
  {H.}~\bibnamefont {Savajols}}, \bibinfo {author} {\bibfnamefont
  {N.}~\bibnamefont {de~S\'er\'eville}}, \bibinfo {author} {\bibfnamefont
  {S.}~\bibnamefont {Sidorchuk}}, \bibinfo {author} {\bibfnamefont
  {D.}~\bibnamefont {Suzuki}}, \bibinfo {author} {\bibfnamefont
  {I.}~\bibnamefont {Strojek}},\ and\ \bibinfo {author} {\bibfnamefont {N.~A.}\
  \bibnamefont {Orr}},\ }\bibfield  {title} {\bibinfo {title} {Structure of
  unbound neutron-rich {$^{9}$He} studied using single-neutron transfer},\
  }\href {https://doi.org/10.1103/PhysRevC.88.034301} {\bibfield  {journal}
  {\bibinfo  {journal} {Phys. Rev. C}\ }\textbf {\bibinfo {volume} {88}},\
  \bibinfo {pages} {034301} (\bibinfo {year} {2013})}\BibitemShut {NoStop}%
\bibitem [{\citenamefont {Bohlen}\ \emph {et~al.}(1988)\citenamefont {Bohlen},
  \citenamefont {Gebauer}, \citenamefont {Kolbert}, \citenamefont {{von
  Oertzen}}, \citenamefont {Stiliaris}, \citenamefont {Wilpert},\ and\
  \citenamefont {Wilpert}}]{Bohlen:1988}%
  \BibitemOpen
  \bibfield  {author} {\bibinfo {author} {\bibfnamefont {H.~G.}\ \bibnamefont
  {Bohlen}}, \bibinfo {author} {\bibfnamefont {B.}~\bibnamefont {Gebauer}},
  \bibinfo {author} {\bibfnamefont {D.}~\bibnamefont {Kolbert}}, \bibinfo
  {author} {\bibfnamefont {W.}~\bibnamefont {{von Oertzen}}}, \bibinfo {author}
  {\bibfnamefont {E.}~\bibnamefont {Stiliaris}}, \bibinfo {author}
  {\bibfnamefont {M.}~\bibnamefont {Wilpert}},\ and\ \bibinfo {author}
  {\bibfnamefont {T.}~\bibnamefont {Wilpert}},\ }\bibfield  {title} {\bibinfo
  {title} {Spectroscopy of {$^9$He} with the {($^{13}$C,$^{13}$O)}-reaction on
  {$^{9}$Be}},\ }\href@noop {} {\bibfield  {journal} {\bibinfo  {journal} {Z.
  Phys. A}\ }\textbf {\bibinfo {volume} {330}},\ \bibinfo {pages} {227}
  (\bibinfo {year} {1988})}\BibitemShut {NoStop}%
\bibitem [{\citenamefont {Votaw}\ \emph {et~al.}(2020)\citenamefont {Votaw},
  \citenamefont {DeYoung}, \citenamefont {Baumann}, \citenamefont {Blake},
  \citenamefont {Boone}, \citenamefont {Brown}, \citenamefont {Chrisman},
  \citenamefont {Finck}, \citenamefont {Frank}, \citenamefont {Gombas},
  \citenamefont {Gu\`eye}, \citenamefont {Hinnefeld}, \citenamefont {Karrick},
  \citenamefont {Kuchera}, \citenamefont {Liu}, \citenamefont {Luther},
  \citenamefont {Ndayisabye}, \citenamefont {Neal}, \citenamefont
  {Owens-Fryar}, \citenamefont {Pereira}, \citenamefont {Persch}, \citenamefont
  {Phan}, \citenamefont {Redpath}, \citenamefont {Rogers}, \citenamefont
  {Stephenson}, \citenamefont {Stiefel}, \citenamefont {Sword}, \citenamefont
  {Wantz},\ and\ \citenamefont {Thoennessen}}]{Votaw:2020}%
  \BibitemOpen
  \bibfield  {author} {\bibinfo {author} {\bibfnamefont {D.}~\bibnamefont
  {Votaw}}, \bibinfo {author} {\bibfnamefont {P.~A.}\ \bibnamefont {DeYoung}},
  \bibinfo {author} {\bibfnamefont {T.}~\bibnamefont {Baumann}}, \bibinfo
  {author} {\bibfnamefont {A.}~\bibnamefont {Blake}}, \bibinfo {author}
  {\bibfnamefont {J.}~\bibnamefont {Boone}}, \bibinfo {author} {\bibfnamefont
  {J.}~\bibnamefont {Brown}}, \bibinfo {author} {\bibfnamefont
  {D.}~\bibnamefont {Chrisman}}, \bibinfo {author} {\bibfnamefont {J.~E.}\
  \bibnamefont {Finck}}, \bibinfo {author} {\bibfnamefont {N.}~\bibnamefont
  {Frank}}, \bibinfo {author} {\bibfnamefont {J.}~\bibnamefont {Gombas}},
  \bibinfo {author} {\bibfnamefont {P.}~\bibnamefont {Gu\`eye}}, \bibinfo
  {author} {\bibfnamefont {J.}~\bibnamefont {Hinnefeld}}, \bibinfo {author}
  {\bibfnamefont {H.}~\bibnamefont {Karrick}}, \bibinfo {author} {\bibfnamefont
  {A.~N.}\ \bibnamefont {Kuchera}}, \bibinfo {author} {\bibfnamefont
  {H.}~\bibnamefont {Liu}}, \bibinfo {author} {\bibfnamefont {B.}~\bibnamefont
  {Luther}}, \bibinfo {author} {\bibfnamefont {F.}~\bibnamefont {Ndayisabye}},
  \bibinfo {author} {\bibfnamefont {M.}~\bibnamefont {Neal}}, \bibinfo {author}
  {\bibfnamefont {J.}~\bibnamefont {Owens-Fryar}}, \bibinfo {author}
  {\bibfnamefont {J.}~\bibnamefont {Pereira}}, \bibinfo {author} {\bibfnamefont
  {C.}~\bibnamefont {Persch}}, \bibinfo {author} {\bibfnamefont
  {T.}~\bibnamefont {Phan}}, \bibinfo {author} {\bibfnamefont {T.}~\bibnamefont
  {Redpath}}, \bibinfo {author} {\bibfnamefont {W.~F.}\ \bibnamefont {Rogers}},
  \bibinfo {author} {\bibfnamefont {S.}~\bibnamefont {Stephenson}}, \bibinfo
  {author} {\bibfnamefont {K.}~\bibnamefont {Stiefel}}, \bibinfo {author}
  {\bibfnamefont {C.}~\bibnamefont {Sword}}, \bibinfo {author} {\bibfnamefont
  {A.}~\bibnamefont {Wantz}},\ and\ \bibinfo {author} {\bibfnamefont
  {M.}~\bibnamefont {Thoennessen}},\ }\bibfield  {title} {\bibinfo {title}
  {Low-lying level structure of the neutron-unbound {$N=7$} isotones},\ }\href
  {https://doi.org/10.1103/PhysRevC.102.014325} {\bibfield  {journal} {\bibinfo
   {journal} {Phys. Rev. C}\ }\textbf {\bibinfo {volume} {102}},\ \bibinfo
  {pages} {014325} (\bibinfo {year} {2020})}\BibitemShut {NoStop}%
\bibitem [{\citenamefont {Chen}\ \emph {et~al.}(2001)\citenamefont {Chen},
  \citenamefont {Blank}, \citenamefont {Brown}, \citenamefont {Chartier},
  \citenamefont {Galonsky}, \citenamefont {Hansen},\ and\ \citenamefont
  {Thoennnessen}}]{Chen:2001}%
  \BibitemOpen
  \bibfield  {author} {\bibinfo {author} {\bibfnamefont {L.}~\bibnamefont
  {Chen}}, \bibinfo {author} {\bibfnamefont {B.}~\bibnamefont {Blank}},
  \bibinfo {author} {\bibfnamefont {B.}~\bibnamefont {Brown}}, \bibinfo
  {author} {\bibfnamefont {M.}~\bibnamefont {Chartier}}, \bibinfo {author}
  {\bibfnamefont {A.}~\bibnamefont {Galonsky}}, \bibinfo {author}
  {\bibfnamefont {P.}~\bibnamefont {Hansen}},\ and\ \bibinfo {author}
  {\bibfnamefont {M.}~\bibnamefont {Thoennnessen}},\ }\bibfield  {title}
  {\bibinfo {title} {Evidence for an l=0 ground state in {$^{9}$He}},\ }\href
  {https://doi.org/https://doi.org/10.1016/S0370-2693(01)00313-6} {\bibfield
  {journal} {\bibinfo  {journal} {Phys. Lett. B}\ }\textbf {\bibinfo {volume}
  {505}},\ \bibinfo {pages} {21} (\bibinfo {year} {2001})}\BibitemShut
  {NoStop}%
\bibitem [{\citenamefont {Golovkov}\ \emph {et~al.}(2007)\citenamefont
  {Golovkov}, \citenamefont {Grigorenko}, \citenamefont {Fomichev},
  \citenamefont {Gorshkov}, \citenamefont {Gorshkov}, \citenamefont {Krupko},
  \citenamefont {Oganessian}, \citenamefont {Rodin}, \citenamefont {Sidorchuk},
  \citenamefont {Slepnev}, \citenamefont {Stepantsov}, \citenamefont
  {Ter-Akopian}, \citenamefont {Wolski}, \citenamefont {Korsheninnikov},
  \citenamefont {Nikolskii}, \citenamefont {Kuzmin}, \citenamefont {Novatskii},
  \citenamefont {Stepanov}, \citenamefont {Roussel-Chomaz},\ and\ \citenamefont
  {Mittig}}]{Golovkov:2007}%
  \BibitemOpen
  \bibfield  {author} {\bibinfo {author} {\bibfnamefont {M.~S.}\ \bibnamefont
  {Golovkov}}, \bibinfo {author} {\bibfnamefont {L.~V.}\ \bibnamefont
  {Grigorenko}}, \bibinfo {author} {\bibfnamefont {A.~S.}\ \bibnamefont
  {Fomichev}}, \bibinfo {author} {\bibfnamefont {A.~V.}\ \bibnamefont
  {Gorshkov}}, \bibinfo {author} {\bibfnamefont {V.~A.}\ \bibnamefont
  {Gorshkov}}, \bibinfo {author} {\bibfnamefont {S.~A.}\ \bibnamefont
  {Krupko}}, \bibinfo {author} {\bibfnamefont {Y.~T.}\ \bibnamefont
  {Oganessian}}, \bibinfo {author} {\bibfnamefont {A.~M.}\ \bibnamefont
  {Rodin}}, \bibinfo {author} {\bibfnamefont {S.~I.}\ \bibnamefont
  {Sidorchuk}}, \bibinfo {author} {\bibfnamefont {R.~S.}\ \bibnamefont
  {Slepnev}}, \bibinfo {author} {\bibfnamefont {S.~V.}\ \bibnamefont
  {Stepantsov}}, \bibinfo {author} {\bibfnamefont {G.~M.}\ \bibnamefont
  {Ter-Akopian}}, \bibinfo {author} {\bibfnamefont {R.}~\bibnamefont {Wolski}},
  \bibinfo {author} {\bibfnamefont {A.~A.}\ \bibnamefont {Korsheninnikov}},
  \bibinfo {author} {\bibfnamefont {E.~Y.}\ \bibnamefont {Nikolskii}}, \bibinfo
  {author} {\bibfnamefont {V.~A.}\ \bibnamefont {Kuzmin}}, \bibinfo {author}
  {\bibfnamefont {B.~G.}\ \bibnamefont {Novatskii}}, \bibinfo {author}
  {\bibfnamefont {D.~N.}\ \bibnamefont {Stepanov}}, \bibinfo {author}
  {\bibfnamefont {P.}~\bibnamefont {Roussel-Chomaz}},\ and\ \bibinfo {author}
  {\bibfnamefont {W.}~\bibnamefont {Mittig}},\ }\bibfield  {title} {\bibinfo
  {title} {New insight into the low-energy $^{9}\mathrm{He}$ spectrum},\ }\href
  {https://doi.org/10.1103/PhysRevC.76.021605} {\bibfield  {journal} {\bibinfo
  {journal} {Phys. Rev. C}\ }\textbf {\bibinfo {volume} {76}},\ \bibinfo
  {pages} {021605(R)} (\bibinfo {year} {2007})}\BibitemShut {NoStop}%
\bibitem [{\citenamefont {Johansson}\ \emph {et~al.}(2010)\citenamefont
  {Johansson}, \citenamefont {Aksyutina}, \citenamefont {Aumann}, \citenamefont
  {Boretzky}, \citenamefont {Borge}, \citenamefont {Chatillon}, \citenamefont
  {Chulkov}, \citenamefont {{Cortina-Gil}}, \citenamefont {{Datta Pramanik}},
  \citenamefont {Emling}, \citenamefont {Forss\'{e}n}, \citenamefont {Fynbo},
  \citenamefont {Geissel}, \citenamefont {Ickert}, \citenamefont {Jonson},
  \citenamefont {Kulessa}, \citenamefont {Langer}, \citenamefont {Lantz},
  \citenamefont {LeBleis}, \citenamefont {Mahata}, \citenamefont {Meister},
  \citenamefont {M\"{u}nzenberg}, \citenamefont {Nilsson}, \citenamefont
  {Nyman}, \citenamefont {Palit}, \citenamefont {Paschalis}, \citenamefont
  {Prokopowicz}, \citenamefont {Reifarth}, \citenamefont {Richter},
  \citenamefont {Riisager}, \citenamefont {Schrieder}, \citenamefont {Simon},
  \citenamefont {S\"{u}mmerer}, \citenamefont {Tengblad}, \citenamefont
  {Weick},\ and\ \citenamefont {Zhukov}}]{Johansson:2010}%
  \BibitemOpen
  \bibfield  {author} {\bibinfo {author} {\bibfnamefont {H.}~\bibnamefont
  {Johansson}}, \bibinfo {author} {\bibfnamefont {Y.}~\bibnamefont
  {Aksyutina}}, \bibinfo {author} {\bibfnamefont {T.}~\bibnamefont {Aumann}},
  \bibinfo {author} {\bibfnamefont {K.}~\bibnamefont {Boretzky}}, \bibinfo
  {author} {\bibfnamefont {M.~J.~G.}\ \bibnamefont {Borge}}, \bibinfo {author}
  {\bibfnamefont {A.}~\bibnamefont {Chatillon}}, \bibinfo {author}
  {\bibfnamefont {L.~V.}\ \bibnamefont {Chulkov}}, \bibinfo {author}
  {\bibfnamefont {D.}~\bibnamefont {{Cortina-Gil}}}, \bibinfo {author}
  {\bibfnamefont {U.}~\bibnamefont {{Datta Pramanik}}}, \bibinfo {author}
  {\bibfnamefont {H.}~\bibnamefont {Emling}}, \bibinfo {author} {\bibfnamefont
  {C.}~\bibnamefont {Forss\'{e}n}}, \bibinfo {author} {\bibfnamefont
  {H.}~\bibnamefont {Fynbo}}, \bibinfo {author} {\bibfnamefont
  {H.}~\bibnamefont {Geissel}}, \bibinfo {author} {\bibfnamefont
  {G.}~\bibnamefont {Ickert}}, \bibinfo {author} {\bibfnamefont
  {B.}~\bibnamefont {Jonson}}, \bibinfo {author} {\bibfnamefont
  {R.}~\bibnamefont {Kulessa}}, \bibinfo {author} {\bibfnamefont
  {C.}~\bibnamefont {Langer}}, \bibinfo {author} {\bibfnamefont
  {M.}~\bibnamefont {Lantz}}, \bibinfo {author} {\bibfnamefont
  {T.}~\bibnamefont {LeBleis}}, \bibinfo {author} {\bibfnamefont
  {K.}~\bibnamefont {Mahata}}, \bibinfo {author} {\bibfnamefont
  {M.}~\bibnamefont {Meister}}, \bibinfo {author} {\bibfnamefont
  {G.}~\bibnamefont {M\"{u}nzenberg}}, \bibinfo {author} {\bibfnamefont
  {T.}~\bibnamefont {Nilsson}}, \bibinfo {author} {\bibfnamefont
  {G.}~\bibnamefont {Nyman}}, \bibinfo {author} {\bibfnamefont
  {R.}~\bibnamefont {Palit}}, \bibinfo {author} {\bibfnamefont
  {S.}~\bibnamefont {Paschalis}}, \bibinfo {author} {\bibfnamefont
  {W.}~\bibnamefont {Prokopowicz}}, \bibinfo {author} {\bibfnamefont
  {R.}~\bibnamefont {Reifarth}}, \bibinfo {author} {\bibfnamefont
  {A.}~\bibnamefont {Richter}}, \bibinfo {author} {\bibfnamefont
  {K.}~\bibnamefont {Riisager}}, \bibinfo {author} {\bibfnamefont
  {G.}~\bibnamefont {Schrieder}}, \bibinfo {author} {\bibfnamefont
  {H.}~\bibnamefont {Simon}}, \bibinfo {author} {\bibfnamefont
  {K.}~\bibnamefont {S\"{u}mmerer}}, \bibinfo {author} {\bibfnamefont
  {O.}~\bibnamefont {Tengblad}}, \bibinfo {author} {\bibfnamefont
  {H.}~\bibnamefont {Weick}},\ and\ \bibinfo {author} {\bibfnamefont
  {M.}~\bibnamefont {Zhukov}},\ }\bibfield  {title} {\bibinfo {title} {The
  unbound isotopes {$^{9,10}$He}},\ }\href
  {https://doi.org/https://doi.org/10.1016/j.nuclphysa.2010.04.006} {\bibfield
  {journal} {\bibinfo  {journal} {Nucl. Phys. A}\ }\textbf {\bibinfo {volume}
  {842}},\ \bibinfo {pages} {15} (\bibinfo {year} {2010})}\BibitemShut
  {NoStop}%
\bibitem [{\citenamefont {Vorabbi}\ \emph {et~al.}(2018)\citenamefont
  {Vorabbi}, \citenamefont {Calci}, \citenamefont {Navr\'atil}, \citenamefont
  {Kruse}, \citenamefont {Quaglioni},\ and\ \citenamefont
  {Hupin}}]{Vorabbi:2018}%
  \BibitemOpen
  \bibfield  {author} {\bibinfo {author} {\bibfnamefont {M.}~\bibnamefont
  {Vorabbi}}, \bibinfo {author} {\bibfnamefont {A.}~\bibnamefont {Calci}},
  \bibinfo {author} {\bibfnamefont {P.}~\bibnamefont {Navr\'atil}}, \bibinfo
  {author} {\bibfnamefont {M.~K.~G.}\ \bibnamefont {Kruse}}, \bibinfo {author}
  {\bibfnamefont {S.}~\bibnamefont {Quaglioni}},\ and\ \bibinfo {author}
  {\bibfnamefont {G.}~\bibnamefont {Hupin}},\ }\bibfield  {title} {\bibinfo
  {title} {Structure of the exotic $^{9}\mathrm{He}$ nucleus from the no-core
  shell model with continuum},\ }\href
  {https://doi.org/10.1103/PhysRevC.97.034314} {\bibfield  {journal} {\bibinfo
  {journal} {Phys. Rev. C}\ }\textbf {\bibinfo {volume} {97}},\ \bibinfo
  {pages} {034314} (\bibinfo {year} {2018})}\BibitemShut {NoStop}%
\bibitem [{\citenamefont {Wallace}\ \emph {et~al.}(2007)\citenamefont
  {Wallace}, \citenamefont {Famiano}, \citenamefont {van Goethem},
  \citenamefont {Rogers}, \citenamefont {Lynch}, \citenamefont {Clifford},
  \citenamefont {Delaunay}, \citenamefont {Lee}, \citenamefont {Labostov},
  \citenamefont {Mocko}, \citenamefont {Morris}, \citenamefont {Moroni},
  \citenamefont {Nett}, \citenamefont {Oostdyk}, \citenamefont {Krishnasamy},
  \citenamefont {Tsang}, \citenamefont {de~Souza}, \citenamefont {Hudan},
  \citenamefont {Sobotka}, \citenamefont {Charity}, \citenamefont {Elson},\
  and\ \citenamefont {Engel}}]{Wallace:2007}%
  \BibitemOpen
  \bibfield  {author} {\bibinfo {author} {\bibfnamefont {M.}~\bibnamefont
  {Wallace}}, \bibinfo {author} {\bibfnamefont {M.}~\bibnamefont {Famiano}},
  \bibinfo {author} {\bibfnamefont {M.-J.}\ \bibnamefont {van Goethem}},
  \bibinfo {author} {\bibfnamefont {A.}~\bibnamefont {Rogers}}, \bibinfo
  {author} {\bibfnamefont {W.}~\bibnamefont {Lynch}}, \bibinfo {author}
  {\bibfnamefont {J.}~\bibnamefont {Clifford}}, \bibinfo {author}
  {\bibfnamefont {F.}~\bibnamefont {Delaunay}}, \bibinfo {author}
  {\bibfnamefont {J.}~\bibnamefont {Lee}}, \bibinfo {author} {\bibfnamefont
  {S.}~\bibnamefont {Labostov}}, \bibinfo {author} {\bibfnamefont
  {M.}~\bibnamefont {Mocko}}, \bibinfo {author} {\bibfnamefont
  {L.}~\bibnamefont {Morris}}, \bibinfo {author} {\bibfnamefont
  {A.}~\bibnamefont {Moroni}}, \bibinfo {author} {\bibfnamefont
  {B.}~\bibnamefont {Nett}}, \bibinfo {author} {\bibfnamefont {D.}~\bibnamefont
  {Oostdyk}}, \bibinfo {author} {\bibfnamefont {R.}~\bibnamefont
  {Krishnasamy}}, \bibinfo {author} {\bibfnamefont {M.}~\bibnamefont {Tsang}},
  \bibinfo {author} {\bibfnamefont {R.}~\bibnamefont {de~Souza}}, \bibinfo
  {author} {\bibfnamefont {S.}~\bibnamefont {Hudan}}, \bibinfo {author}
  {\bibfnamefont {L.}~\bibnamefont {Sobotka}}, \bibinfo {author} {\bibfnamefont
  {R.}~\bibnamefont {Charity}}, \bibinfo {author} {\bibfnamefont
  {J.}~\bibnamefont {Elson}},\ and\ \bibinfo {author} {\bibfnamefont
  {G.}~\bibnamefont {Engel}},\ }\bibfield  {title} {\bibinfo {title} {The high
  resolution array ({HiRA}) for rare isotope beam experiments},\ }\href
  {https://doi.org/http://dx.doi.org/10.1016/j.nima.2007.08.248} {\bibfield
  {journal} {\bibinfo  {journal} {Nucl. Instrum. Methods Phys. Res. A}\
  }\textbf {\bibinfo {volume} {583}},\ \bibinfo {pages} {302 } (\bibinfo {year}
  {2007})}\BibitemShut {NoStop}%
\bibitem [{sup()}]{sup}%
  \BibitemOpen
  \href@noop {} {}\bibinfo {note} {See Supplemental Material at [URL inserted
  by publisher] for more details on beam production, experimental apparatus,
  Monte Carlo simulations, analysis of background fluctuations, theory
  calculations, and the list of the predicted states; it includes
  Refs.~\cite{Bazin:2009,Charity:2019,Lane:1958,Charity:2010,Brown:2014,Egorova:2012,Charity:2011,Wylie:2021,Michel2006a,Jaganathen:2017,Fossez:2018,Wang:2019,ENSDF,Kalanee:2013,Cowan:1998,Rolke:2003,Pekka:2002,ATLAS,CMS,Lyons:2013}}\BibitemShut
  {NoStop}%
\bibitem [{\citenamefont {Webb}\ \emph
  {et~al.}(2019{\natexlab{a}})\citenamefont {Webb}, \citenamefont {Charity},
  \citenamefont {Elson}, \citenamefont {Hoff}, \citenamefont {Pruitt},
  \citenamefont {Sobotka}, \citenamefont {Brown}, \citenamefont {Barney},
  \citenamefont {Cerizza}, \citenamefont {Estee}, \citenamefont {Jhang},
  \citenamefont {Lynch}, \citenamefont {Manfredi}, \citenamefont {Morfouace},
  \citenamefont {Santamaria}, \citenamefont {Sweany}, \citenamefont {Tsang},
  \citenamefont {Tsang}, \citenamefont {Wang}, \citenamefont {Zhang},
  \citenamefont {Zhu}, \citenamefont {Kuvin}, \citenamefont {McNeel},
  \citenamefont {Smith}, \citenamefont {Wuosmaa},\ and\ \citenamefont
  {Chajecki}}]{Webb:2019a}%
  \BibitemOpen
  \bibfield  {author} {\bibinfo {author} {\bibfnamefont {T.~B.}\ \bibnamefont
  {Webb}}, \bibinfo {author} {\bibfnamefont {R.~J.}\ \bibnamefont {Charity}},
  \bibinfo {author} {\bibfnamefont {J.~M.}\ \bibnamefont {Elson}}, \bibinfo
  {author} {\bibfnamefont {D.~E.~M.}\ \bibnamefont {Hoff}}, \bibinfo {author}
  {\bibfnamefont {C.~D.}\ \bibnamefont {Pruitt}}, \bibinfo {author}
  {\bibfnamefont {L.~G.}\ \bibnamefont {Sobotka}}, \bibinfo {author}
  {\bibfnamefont {K.~W.}\ \bibnamefont {Brown}}, \bibinfo {author}
  {\bibfnamefont {J.}~\bibnamefont {Barney}}, \bibinfo {author} {\bibfnamefont
  {G.}~\bibnamefont {Cerizza}}, \bibinfo {author} {\bibfnamefont
  {J.}~\bibnamefont {Estee}}, \bibinfo {author} {\bibfnamefont
  {G.}~\bibnamefont {Jhang}}, \bibinfo {author} {\bibfnamefont {W.~G.}\
  \bibnamefont {Lynch}}, \bibinfo {author} {\bibfnamefont {J.}~\bibnamefont
  {Manfredi}}, \bibinfo {author} {\bibfnamefont {P.}~\bibnamefont {Morfouace}},
  \bibinfo {author} {\bibfnamefont {C.}~\bibnamefont {Santamaria}}, \bibinfo
  {author} {\bibfnamefont {S.}~\bibnamefont {Sweany}}, \bibinfo {author}
  {\bibfnamefont {M.~B.}\ \bibnamefont {Tsang}}, \bibinfo {author}
  {\bibfnamefont {T.}~\bibnamefont {Tsang}}, \bibinfo {author} {\bibfnamefont
  {S.~M.}\ \bibnamefont {Wang}}, \bibinfo {author} {\bibfnamefont
  {Y.}~\bibnamefont {Zhang}}, \bibinfo {author} {\bibfnamefont
  {K.}~\bibnamefont {Zhu}}, \bibinfo {author} {\bibfnamefont {S.~A.}\
  \bibnamefont {Kuvin}}, \bibinfo {author} {\bibfnamefont {D.}~\bibnamefont
  {McNeel}}, \bibinfo {author} {\bibfnamefont {J.}~\bibnamefont {Smith}},
  \bibinfo {author} {\bibfnamefont {A.~H.}\ \bibnamefont {Wuosmaa}},\ and\
  \bibinfo {author} {\bibfnamefont {Z.}~\bibnamefont {Chajecki}},\ }\bibfield
  {title} {\bibinfo {title} {Particle decays of levels in $^{11,12}\mathrm{N}$
  and $^{12}\mathrm{O}$ investigated with the invariant-mass method},\ }\href
  {https://doi.org/10.1103/PhysRevC.100.024306} {\bibfield  {journal} {\bibinfo
   {journal} {Phys. Rev. C}\ }\textbf {\bibinfo {volume} {100}},\ \bibinfo
  {pages} {024306} (\bibinfo {year} {2019}{\natexlab{a}})}\BibitemShut
  {NoStop}%
\bibitem [{\citenamefont {Webb}\ \emph {et~al.}(2020)\citenamefont {Webb},
  \citenamefont {Charity}, \citenamefont {Elson}, \citenamefont {Hoff},
  \citenamefont {Pruitt}, \citenamefont {Sobotka}, \citenamefont {Brown},
  \citenamefont {Barney}, \citenamefont {Cerizza}, \citenamefont {Estee},
  \citenamefont {Lynch}, \citenamefont {Manfredi}, \citenamefont {Morfouace},
  \citenamefont {Santamaria}, \citenamefont {Sweany}, \citenamefont {Tsang},
  \citenamefont {Tsang}, \citenamefont {Zhang}, \citenamefont {Zhu},
  \citenamefont {Kuvin}, \citenamefont {McNeel}, \citenamefont {Smith},
  \citenamefont {Wuosmaa},\ and\ \citenamefont {Chajecki}}]{Webb:2020}%
  \BibitemOpen
  \bibfield  {author} {\bibinfo {author} {\bibfnamefont {T.~B.}\ \bibnamefont
  {Webb}}, \bibinfo {author} {\bibfnamefont {R.~J.}\ \bibnamefont {Charity}},
  \bibinfo {author} {\bibfnamefont {J.~M.}\ \bibnamefont {Elson}}, \bibinfo
  {author} {\bibfnamefont {D.~E.~M.}\ \bibnamefont {Hoff}}, \bibinfo {author}
  {\bibfnamefont {C.~D.}\ \bibnamefont {Pruitt}}, \bibinfo {author}
  {\bibfnamefont {L.~G.}\ \bibnamefont {Sobotka}}, \bibinfo {author}
  {\bibfnamefont {K.~W.}\ \bibnamefont {Brown}}, \bibinfo {author}
  {\bibfnamefont {J.}~\bibnamefont {Barney}}, \bibinfo {author} {\bibfnamefont
  {G.}~\bibnamefont {Cerizza}}, \bibinfo {author} {\bibfnamefont
  {J.}~\bibnamefont {Estee}}, \bibinfo {author} {\bibfnamefont {W.~G.}\
  \bibnamefont {Lynch}}, \bibinfo {author} {\bibfnamefont {J.}~\bibnamefont
  {Manfredi}}, \bibinfo {author} {\bibfnamefont {P.}~\bibnamefont {Morfouace}},
  \bibinfo {author} {\bibfnamefont {C.}~\bibnamefont {Santamaria}}, \bibinfo
  {author} {\bibfnamefont {S.}~\bibnamefont {Sweany}}, \bibinfo {author}
  {\bibfnamefont {M.~B.}\ \bibnamefont {Tsang}}, \bibinfo {author}
  {\bibfnamefont {T.}~\bibnamefont {Tsang}}, \bibinfo {author} {\bibfnamefont
  {Y.}~\bibnamefont {Zhang}}, \bibinfo {author} {\bibfnamefont
  {K.}~\bibnamefont {Zhu}}, \bibinfo {author} {\bibfnamefont {S.~A.}\
  \bibnamefont {Kuvin}}, \bibinfo {author} {\bibfnamefont {D.}~\bibnamefont
  {McNeel}}, \bibinfo {author} {\bibfnamefont {J.}~\bibnamefont {Smith}},
  \bibinfo {author} {\bibfnamefont {A.~H.}\ \bibnamefont {Wuosmaa}},\ and\
  \bibinfo {author} {\bibfnamefont {Z.}~\bibnamefont {Chajecki}},\ }\bibfield
  {title} {\bibinfo {title} {Invariant-mass spectrum of $^{11}\mathrm{O}$},\
  }\href {https://doi.org/10.1103/PhysRevC.101.044317} {\bibfield  {journal}
  {\bibinfo  {journal} {Phys. Rev. C}\ }\textbf {\bibinfo {volume} {101}},\
  \bibinfo {pages} {044317} (\bibinfo {year} {2020})}\BibitemShut {NoStop}%
\bibitem [{\citenamefont {Webb}\ \emph
  {et~al.}(2019{\natexlab{b}})\citenamefont {Webb}, \citenamefont {Wang},
  \citenamefont {Brown}, \citenamefont {Charity}, \citenamefont {Elson},
  \citenamefont {Barney}, \citenamefont {Cerizza}, \citenamefont {Chajecki},
  \citenamefont {Estee}, \citenamefont {Hoff}, \citenamefont {Kuvin},
  \citenamefont {Lynch}, \citenamefont {Manfredi}, \citenamefont {McNeel},
  \citenamefont {Morfouace}, \citenamefont {Nazarewicz}, \citenamefont
  {Pruitt}, \citenamefont {Santamaria}, \citenamefont {Smith}, \citenamefont
  {Sobotka}, \citenamefont {Sweany}, \citenamefont {Tsang}, \citenamefont
  {Tsang}, \citenamefont {Wuosmaa}, \citenamefont {Zhang},\ and\ \citenamefont
  {Zhu}}]{Webb:2019}%
  \BibitemOpen
  \bibfield  {author} {\bibinfo {author} {\bibfnamefont {T.~B.}\ \bibnamefont
  {Webb}}, \bibinfo {author} {\bibfnamefont {S.~M.}\ \bibnamefont {Wang}},
  \bibinfo {author} {\bibfnamefont {K.~W.}\ \bibnamefont {Brown}}, \bibinfo
  {author} {\bibfnamefont {R.~J.}\ \bibnamefont {Charity}}, \bibinfo {author}
  {\bibfnamefont {J.~M.}\ \bibnamefont {Elson}}, \bibinfo {author}
  {\bibfnamefont {J.}~\bibnamefont {Barney}}, \bibinfo {author} {\bibfnamefont
  {G.}~\bibnamefont {Cerizza}}, \bibinfo {author} {\bibfnamefont
  {Z.}~\bibnamefont {Chajecki}}, \bibinfo {author} {\bibfnamefont
  {J.}~\bibnamefont {Estee}}, \bibinfo {author} {\bibfnamefont {D.~E.~M.}\
  \bibnamefont {Hoff}}, \bibinfo {author} {\bibfnamefont {S.~A.}\ \bibnamefont
  {Kuvin}}, \bibinfo {author} {\bibfnamefont {W.~G.}\ \bibnamefont {Lynch}},
  \bibinfo {author} {\bibfnamefont {J.}~\bibnamefont {Manfredi}}, \bibinfo
  {author} {\bibfnamefont {D.}~\bibnamefont {McNeel}}, \bibinfo {author}
  {\bibfnamefont {P.}~\bibnamefont {Morfouace}}, \bibinfo {author}
  {\bibfnamefont {W.}~\bibnamefont {Nazarewicz}}, \bibinfo {author}
  {\bibfnamefont {C.~D.}\ \bibnamefont {Pruitt}}, \bibinfo {author}
  {\bibfnamefont {C.}~\bibnamefont {Santamaria}}, \bibinfo {author}
  {\bibfnamefont {J.}~\bibnamefont {Smith}}, \bibinfo {author} {\bibfnamefont
  {L.~G.}\ \bibnamefont {Sobotka}}, \bibinfo {author} {\bibfnamefont
  {S.}~\bibnamefont {Sweany}}, \bibinfo {author} {\bibfnamefont {C.~Y.}\
  \bibnamefont {Tsang}}, \bibinfo {author} {\bibfnamefont {M.~B.}\ \bibnamefont
  {Tsang}}, \bibinfo {author} {\bibfnamefont {A.~H.}\ \bibnamefont {Wuosmaa}},
  \bibinfo {author} {\bibfnamefont {Y.}~\bibnamefont {Zhang}},\ and\ \bibinfo
  {author} {\bibfnamefont {K.}~\bibnamefont {Zhu}},\ }\bibfield  {title}
  {\bibinfo {title} {First observation of unbound $^{11}\mathrm{O}$, the mirror
  of the halo nucleus $^{11}\mathrm{Li}$},\ }\href
  {https://doi.org/10.1103/PhysRevLett.122.122501} {\bibfield  {journal}
  {\bibinfo  {journal} {Phys. Rev. Lett.}\ }\textbf {\bibinfo {volume} {122}},\
  \bibinfo {pages} {122501} (\bibinfo {year} {2019}{\natexlab{b}})}\BibitemShut
  {NoStop}%
\bibitem [{\citenamefont {Charity}\ \emph
  {et~al.}(2021{\natexlab{c}})\citenamefont {Charity}, \citenamefont {Webb},
  \citenamefont {Elson}, \citenamefont {Hoff}, \citenamefont {Pruitt},
  \citenamefont {Sobotka}, \citenamefont {Navr\'atil}, \citenamefont {Hupin},
  \citenamefont {Kravvaris}, \citenamefont {Quaglioni}, \citenamefont {Brown},
  \citenamefont {Cerizza}, \citenamefont {Estee}, \citenamefont {Lynch},
  \citenamefont {Manfredi}, \citenamefont {Morfouace}, \citenamefont
  {Santamaria}, \citenamefont {Sweany}, \citenamefont {Tsang}, \citenamefont
  {Tsang}, \citenamefont {Zhu}, \citenamefont {Kuvin}, \citenamefont {McNeel},
  \citenamefont {Smith}, \citenamefont {Wuosmaa},\ and\ \citenamefont
  {Chajecki}}]{Charity:2021a}%
  \BibitemOpen
  \bibfield  {author} {\bibinfo {author} {\bibfnamefont {R.~J.}\ \bibnamefont
  {Charity}}, \bibinfo {author} {\bibfnamefont {T.~B.}\ \bibnamefont {Webb}},
  \bibinfo {author} {\bibfnamefont {J.~M.}\ \bibnamefont {Elson}}, \bibinfo
  {author} {\bibfnamefont {D.~E.~M.}\ \bibnamefont {Hoff}}, \bibinfo {author}
  {\bibfnamefont {C.~D.}\ \bibnamefont {Pruitt}}, \bibinfo {author}
  {\bibfnamefont {L.~G.}\ \bibnamefont {Sobotka}}, \bibinfo {author}
  {\bibfnamefont {P.}~\bibnamefont {Navr\'atil}}, \bibinfo {author}
  {\bibfnamefont {G.}~\bibnamefont {Hupin}}, \bibinfo {author} {\bibfnamefont
  {K.}~\bibnamefont {Kravvaris}}, \bibinfo {author} {\bibfnamefont
  {S.}~\bibnamefont {Quaglioni}}, \bibinfo {author} {\bibfnamefont {K.~W.}\
  \bibnamefont {Brown}}, \bibinfo {author} {\bibfnamefont {G.}~\bibnamefont
  {Cerizza}}, \bibinfo {author} {\bibfnamefont {J.}~\bibnamefont {Estee}},
  \bibinfo {author} {\bibfnamefont {W.~G.}\ \bibnamefont {Lynch}}, \bibinfo
  {author} {\bibfnamefont {J.}~\bibnamefont {Manfredi}}, \bibinfo {author}
  {\bibfnamefont {P.}~\bibnamefont {Morfouace}}, \bibinfo {author}
  {\bibfnamefont {C.}~\bibnamefont {Santamaria}}, \bibinfo {author}
  {\bibfnamefont {S.}~\bibnamefont {Sweany}}, \bibinfo {author} {\bibfnamefont
  {M.~B.}\ \bibnamefont {Tsang}}, \bibinfo {author} {\bibfnamefont
  {T.}~\bibnamefont {Tsang}}, \bibinfo {author} {\bibfnamefont
  {K.}~\bibnamefont {Zhu}}, \bibinfo {author} {\bibfnamefont {S.~A.}\
  \bibnamefont {Kuvin}}, \bibinfo {author} {\bibfnamefont {D.}~\bibnamefont
  {McNeel}}, \bibinfo {author} {\bibfnamefont {J.}~\bibnamefont {Smith}},
  \bibinfo {author} {\bibfnamefont {A.~H.}\ \bibnamefont {Wuosmaa}},\ and\
  \bibinfo {author} {\bibfnamefont {Z.}~\bibnamefont {Chajecki}},\ }\bibfield
  {title} {\bibinfo {title} {Using spin alignment of inelastically excited
  nuclei in fast beams to assign spins: The spectroscopy of $^{13}\mathrm{O}$
  as a test case},\ }\href {https://doi.org/10.1103/PhysRevC.104.024325}
  {\bibfield  {journal} {\bibinfo  {journal} {Phys. Rev. C}\ }\textbf {\bibinfo
  {volume} {104}},\ \bibinfo {pages} {024325} (\bibinfo {year}
  {2021}{\natexlab{c}})}\BibitemShut {NoStop}%
\bibitem [{\citenamefont {Charity}\ and\ \citenamefont
  {Sobotka}(2023)}]{Charity:2023}%
  \BibitemOpen
  \bibfield  {author} {\bibinfo {author} {\bibfnamefont {R.~J.}\ \bibnamefont
  {Charity}}\ and\ \bibinfo {author} {\bibfnamefont {L.~G.}\ \bibnamefont
  {Sobotka}},\ }\bibfield  {title} {\bibinfo {title} {Invariant-mass
  spectroscopy in projectile-fragmentation reactions},\ }\href@noop {}
  {\bibfield  {journal} {\bibinfo  {journal} {Phys. Rev. C companion paper}\ }
  (\bibinfo {year} {2023})}\BibitemShut {NoStop}%
\bibitem [{\citenamefont {Michel}\ \emph {et~al.}(2002)\citenamefont {Michel},
  \citenamefont {Nazarewicz}, \citenamefont {P\l{}oszajczak},\ and\
  \citenamefont {Bennaceur}}]{Michel:2002}%
  \BibitemOpen
  \bibfield  {author} {\bibinfo {author} {\bibfnamefont {N.}~\bibnamefont
  {Michel}}, \bibinfo {author} {\bibfnamefont {W.}~\bibnamefont {Nazarewicz}},
  \bibinfo {author} {\bibfnamefont {M.}~\bibnamefont {P\l{}oszajczak}},\ and\
  \bibinfo {author} {\bibfnamefont {K.}~\bibnamefont {Bennaceur}},\ }\bibfield
  {title} {\bibinfo {title} {Gamow shell model description of neutron-rich
  nuclei},\ }\href {https://doi.org/10.1103/PhysRevLett.89.042502} {\bibfield
  {journal} {\bibinfo  {journal} {Phys. Rev. Lett.}\ }\textbf {\bibinfo
  {volume} {89}},\ \bibinfo {pages} {042502} (\bibinfo {year}
  {2002})}\BibitemShut {NoStop}%
\bibitem [{\citenamefont {Michel}\ \emph {et~al.}(2009)\citenamefont {Michel},
  \citenamefont {Nazarewicz}, \citenamefont {P{\l}oszajczak},\ and\
  \citenamefont {Vertse}}]{Michel2009}%
  \BibitemOpen
  \bibfield  {author} {\bibinfo {author} {\bibfnamefont {N.}~\bibnamefont
  {Michel}}, \bibinfo {author} {\bibfnamefont {W.}~\bibnamefont {Nazarewicz}},
  \bibinfo {author} {\bibfnamefont {M.}~\bibnamefont {P{\l}oszajczak}},\ and\
  \bibinfo {author} {\bibfnamefont {T.}~\bibnamefont {Vertse}},\ }\bibfield
  {title} {\bibinfo {title} {{Shell model in the complex energy plane}},\
  }\href {https://doi.org/10.1088/0954-3899/36/1/013101} {\bibfield  {journal}
  {\bibinfo  {journal} {J. Phys. G}\ }\textbf {\bibinfo {volume} {36}},\
  \bibinfo {pages} {40} (\bibinfo {year} {2009})}\BibitemShut {NoStop}%
\bibitem [{\citenamefont {Michel}\ and\ \citenamefont
  {P{\l}oszajczak}(2021)}]{Michel:2021}%
  \BibitemOpen
  \bibfield  {author} {\bibinfo {author} {\bibfnamefont {N.}~\bibnamefont
  {Michel}}\ and\ \bibinfo {author} {\bibfnamefont {M.}~\bibnamefont
  {P{\l}oszajczak}},\ }\href
  {https://doi.org/https://doi.org/10.1007/978-3-030-69356-5} {\emph {\bibinfo
  {title} {Gamow Shell Model, The Unified Theory of Nuclear Structure and
  Reactions}}},\ \bibinfo {series} {Lecture Notes in Physics}, Vol.\ \bibinfo
  {volume} {983}\ (\bibinfo  {publisher} {Springer, Cham},\ \bibinfo {year}
  {2021})\BibitemShut {NoStop}%
\bibitem [{\citenamefont {Wylie}\ \emph {et~al.}(2021)\citenamefont {Wylie},
  \citenamefont {Oko{\l}owicz}, \citenamefont {Nazarewicz}, \citenamefont
  {P{\l}oszajczak}, \citenamefont {Wang}, \citenamefont {Mao},\ and\
  \citenamefont {Michel}}]{Wylie:2021}%
  \BibitemOpen
  \bibfield  {author} {\bibinfo {author} {\bibfnamefont {J.}~\bibnamefont
  {Wylie}}, \bibinfo {author} {\bibfnamefont {J.}~\bibnamefont {Oko{\l}owicz}},
  \bibinfo {author} {\bibfnamefont {W.}~\bibnamefont {Nazarewicz}}, \bibinfo
  {author} {\bibfnamefont {M.}~\bibnamefont {P{\l}oszajczak}}, \bibinfo
  {author} {\bibfnamefont {S.~M.}\ \bibnamefont {Wang}}, \bibinfo {author}
  {\bibfnamefont {X.}~\bibnamefont {Mao}},\ and\ \bibinfo {author}
  {\bibfnamefont {N.}~\bibnamefont {Michel}},\ }\bibfield  {title} {\bibinfo
  {title} {Spectroscopic factors in dripline nuclei},\ }\href@noop {}
  {\bibfield  {journal} {\bibinfo  {journal} {Phys. Rev. C}\ }\textbf {\bibinfo
  {volume} {104}},\ \bibinfo {pages} {L061301} (\bibinfo {year}
  {2021})}\BibitemShut {NoStop}%
\bibitem [{\citenamefont {Berggren}(1968)}]{Berggren:1968}%
  \BibitemOpen
  \bibfield  {author} {\bibinfo {author} {\bibfnamefont {T.}~\bibnamefont
  {Berggren}},\ }\bibfield  {title} {\bibinfo {title} {On the use of resonant
  states in eigenfunction expansions of scattering and reaction amplitudes},\
  }\href {https://doi.org/10.1016/0375-9474(68)90593-9} {\bibfield  {journal}
  {\bibinfo  {journal} {Nucl. Phys. A}\ }\textbf {\bibinfo {volume} {109}},\
  \bibinfo {pages} {265} (\bibinfo {year} {1968})}\BibitemShut {NoStop}%
\bibitem [{\citenamefont {Jaganathen}\ \emph {et~al.}(2017)\citenamefont
  {Jaganathen}, \citenamefont {{{Id} Betan}}, \citenamefont {Michel},
  \citenamefont {Nazarewicz},\ and\ \citenamefont
  {P\l{}oszajczak}}]{Jaganathen:2017}%
  \BibitemOpen
  \bibfield  {author} {\bibinfo {author} {\bibfnamefont {Y.}~\bibnamefont
  {Jaganathen}}, \bibinfo {author} {\bibfnamefont {R.~M.}\ \bibnamefont {{{Id}
  Betan}}}, \bibinfo {author} {\bibfnamefont {N.}~\bibnamefont {Michel}},
  \bibinfo {author} {\bibfnamefont {W.}~\bibnamefont {Nazarewicz}},\ and\
  \bibinfo {author} {\bibfnamefont {M.}~\bibnamefont {P\l{}oszajczak}},\
  }\bibfield  {title} {\bibinfo {title} {Quantified {Gamow} shell model
  interaction for $psd$-shell nuclei},\ }\href
  {https://doi.org/10.1103/PhysRevC.96.054316} {\bibfield  {journal} {\bibinfo
  {journal} {Phys. Rev. C}\ }\textbf {\bibinfo {volume} {96}},\ \bibinfo
  {pages} {054316} (\bibinfo {year} {2017})}\BibitemShut {NoStop}%
\bibitem [{\citenamefont {Mao}\ \emph {et~al.}(2020)\citenamefont {Mao},
  \citenamefont {Rotureau}, \citenamefont {Nazarewicz}, \citenamefont {Michel},
  \citenamefont {{{Id} Betan}},\ and\ \citenamefont {Jaganathen}}]{Mao:2020}%
  \BibitemOpen
  \bibfield  {author} {\bibinfo {author} {\bibfnamefont {X.}~\bibnamefont
  {Mao}}, \bibinfo {author} {\bibfnamefont {J.}~\bibnamefont {Rotureau}},
  \bibinfo {author} {\bibfnamefont {W.}~\bibnamefont {Nazarewicz}}, \bibinfo
  {author} {\bibfnamefont {N.}~\bibnamefont {Michel}}, \bibinfo {author}
  {\bibfnamefont {R.~M.}\ \bibnamefont {{{Id} Betan}}},\ and\ \bibinfo {author}
  {\bibfnamefont {Y.}~\bibnamefont {Jaganathen}},\ }\bibfield  {title}
  {\bibinfo {title} {{Gamow-shell-model description of Li isotopes and their
  mirror partners}},\ }\href {https://doi.org/10.1103/PhysRevC.102.024309}
  {\bibfield  {journal} {\bibinfo  {journal} {Phys. Rev. C}\ }\textbf {\bibinfo
  {volume} {102}},\ \bibinfo {pages} {024309} (\bibinfo {year}
  {2020})}\BibitemShut {NoStop}%
\bibitem [{\citenamefont {Myo}\ \emph {et~al.}(2021)\citenamefont {Myo},
  \citenamefont {Odsuren},\ and\ \citenamefont {Kat\={o}}}]{Myo:2021}%
  \BibitemOpen
  \bibfield  {author} {\bibinfo {author} {\bibfnamefont {T.}~\bibnamefont
  {Myo}}, \bibinfo {author} {\bibfnamefont {M.}~\bibnamefont {Odsuren}},\ and\
  \bibinfo {author} {\bibfnamefont {K.}~\bibnamefont {Kat\={o}}},\ }\bibfield
  {title} {\bibinfo {title} {Five-body resonances in $^{8}\mathrm{He}$ and
  $^{8}\mathrm{C}$ using the complex scaling method},\ }\href
  {https://doi.org/10.1103/PhysRevC.104.044306} {\bibfield  {journal} {\bibinfo
   {journal} {Phys. Rev. C}\ }\textbf {\bibinfo {volume} {104}},\ \bibinfo
  {pages} {044306} (\bibinfo {year} {2021})}\BibitemShut {NoStop}%
\bibitem [{\citenamefont {Wang}\ \emph {et~al.}(2019)\citenamefont {Wang},
  \citenamefont {Nazarewicz}, \citenamefont {Charity},\ and\ \citenamefont
  {Sobotka}}]{Wang:2019}%
  \BibitemOpen
  \bibfield  {author} {\bibinfo {author} {\bibfnamefont {S.~M.}\ \bibnamefont
  {Wang}}, \bibinfo {author} {\bibfnamefont {W.}~\bibnamefont {Nazarewicz}},
  \bibinfo {author} {\bibfnamefont {R.~J.}\ \bibnamefont {Charity}},\ and\
  \bibinfo {author} {\bibfnamefont {L.~G.}\ \bibnamefont {Sobotka}},\
  }\bibfield  {title} {\bibinfo {title} {Structure and decay of the extremely
  proton-rich nuclei $^{11,12}\mathrm{O}$},\ }\href
  {https://doi.org/10.1103/PhysRevC.99.054302} {\bibfield  {journal} {\bibinfo
  {journal} {Phys. Rev. C}\ }\textbf {\bibinfo {volume} {99}},\ \bibinfo
  {pages} {054302} (\bibinfo {year} {2019})}\BibitemShut {NoStop}%
\bibitem [{\citenamefont {Wang}\ and\ \citenamefont
  {Nazarewicz}(2021)}]{Wang:2021}%
  \BibitemOpen
  \bibfield  {author} {\bibinfo {author} {\bibfnamefont {S.~M.}\ \bibnamefont
  {Wang}}\ and\ \bibinfo {author} {\bibfnamefont {W.}~\bibnamefont
  {Nazarewicz}},\ }\bibfield  {title} {\bibinfo {title} {Fermion pair dynamics
  in open quantum systems},\ }\href
  {https://doi.org/10.1103/PhysRevLett.126.142501} {\bibfield  {journal}
  {\bibinfo  {journal} {Phys. Rev. Lett.}\ }\textbf {\bibinfo {volume} {126}},\
  \bibinfo {pages} {142501} (\bibinfo {year} {2021})}\BibitemShut {NoStop}%
\bibitem [{\citenamefont {Lane}\ and\ \citenamefont
  {Thomas}(1958)}]{Lane:1958}%
  \BibitemOpen
  \bibfield  {author} {\bibinfo {author} {\bibfnamefont {A.~M.}\ \bibnamefont
  {Lane}}\ and\ \bibinfo {author} {\bibfnamefont {R.~G.}\ \bibnamefont
  {Thomas}},\ }\bibfield  {title} {\bibinfo {title} {{$R$-matrix} theory of
  nuclear reactions},\ }\href {https://doi.org/10.1103/RevModPhys.30.257}
  {\bibfield  {journal} {\bibinfo  {journal} {Rev. Mod. Phys.}\ }\textbf
  {\bibinfo {volume} {30}},\ \bibinfo {pages} {257} (\bibinfo {year}
  {1958})}\BibitemShut {NoStop}%
\bibitem [{\citenamefont {Fossez}\ \emph {et~al.}(2018)\citenamefont {Fossez},
  \citenamefont {Rotureau},\ and\ \citenamefont {Nazarewicz}}]{Fossez:2018}%
  \BibitemOpen
  \bibfield  {author} {\bibinfo {author} {\bibfnamefont {K.}~\bibnamefont
  {Fossez}}, \bibinfo {author} {\bibfnamefont {J.}~\bibnamefont {Rotureau}},\
  and\ \bibinfo {author} {\bibfnamefont {W.}~\bibnamefont {Nazarewicz}},\
  }\bibfield  {title} {\bibinfo {title} {Energy spectrum of neutron-rich helium
  isotopes: Complex made simple},\ }\href
  {https://doi.org/10.1103/PhysRevC.98.061302} {\bibfield  {journal} {\bibinfo
  {journal} {Phys. Rev. C}\ }\textbf {\bibinfo {volume} {98}},\ \bibinfo
  {pages} {061302(R)} (\bibinfo {year} {2018})}\BibitemShut {NoStop}%
\bibitem [{\citenamefont {Michel}\ \emph {et~al.}(2006)\citenamefont {Michel},
  \citenamefont {Nazarewicz}, \citenamefont {P\l{}oszajczak},\ and\
  \citenamefont {Rotureau}}]{Michel2006a}%
  \BibitemOpen
  \bibfield  {author} {\bibinfo {author} {\bibfnamefont {N.}~\bibnamefont
  {Michel}}, \bibinfo {author} {\bibfnamefont {W.}~\bibnamefont {Nazarewicz}},
  \bibinfo {author} {\bibfnamefont {M.}~\bibnamefont {P\l{}oszajczak}},\ and\
  \bibinfo {author} {\bibfnamefont {J.}~\bibnamefont {Rotureau}},\ }\bibfield
  {title} {\bibinfo {title} {Antibound states and halo formation in the {Gamow}
  shell model},\ }\href {https://doi.org/10.1103/PhysRevC.74.054305} {\bibfield
   {journal} {\bibinfo  {journal} {Phys. Rev. C}\ }\textbf {\bibinfo {volume}
  {74}},\ \bibinfo {pages} {054305} (\bibinfo {year} {2006})}\BibitemShut
  {NoStop}%
\bibitem [{\citenamefont {Bazin}\ \emph {et~al.}(2009)\citenamefont {Bazin},
  \citenamefont {Andreev}, \citenamefont {Becerril}, \citenamefont
  {Dol\'{e}ans}, \citenamefont {Mantica}, \citenamefont {Ottarson},
  \citenamefont {Schatz}, \citenamefont {Stoker},\ and\ \citenamefont
  {Vincent}}]{Bazin:2009}%
  \BibitemOpen
  \bibfield  {author} {\bibinfo {author} {\bibfnamefont {D.}~\bibnamefont
  {Bazin}}, \bibinfo {author} {\bibfnamefont {V.}~\bibnamefont {Andreev}},
  \bibinfo {author} {\bibfnamefont {A.}~\bibnamefont {Becerril}}, \bibinfo
  {author} {\bibfnamefont {M.}~\bibnamefont {Dol\'{e}ans}}, \bibinfo {author}
  {\bibfnamefont {P.}~\bibnamefont {Mantica}}, \bibinfo {author} {\bibfnamefont
  {J.}~\bibnamefont {Ottarson}}, \bibinfo {author} {\bibfnamefont
  {H.}~\bibnamefont {Schatz}}, \bibinfo {author} {\bibfnamefont
  {J.}~\bibnamefont {Stoker}},\ and\ \bibinfo {author} {\bibfnamefont
  {J.}~\bibnamefont {Vincent}},\ }\bibfield  {title} {\bibinfo {title} {Radio
  frequency fragment separator at {NSCL}},\ }\href
  {https://doi.org/https://doi.org/10.1016/j.nima.2009.05.100} {\bibfield
  {journal} {\bibinfo  {journal} {Nucl. Instrum. Methods A}\ }\textbf {\bibinfo
  {volume} {606}},\ \bibinfo {pages} {314 } (\bibinfo {year}
  {2009})}\BibitemShut {NoStop}%
\bibitem [{\citenamefont {Charity}\ \emph {et~al.}(2019)\citenamefont
  {Charity}, \citenamefont {Brown}, \citenamefont {Elson}, \citenamefont
  {Reviol}, \citenamefont {Sobotka}, \citenamefont {Buhro}, \citenamefont
  {Chajecki}, \citenamefont {Lynch}, \citenamefont {Manfredi}, \citenamefont
  {Shane}, \citenamefont {Showalter}, \citenamefont {Tsang}, \citenamefont
  {Weisshaar}, \citenamefont {Winkelbauer}, \citenamefont {Bedoor},
  \citenamefont {McNeel},\ and\ \citenamefont {Wuosmaa}}]{Charity:2019}%
  \BibitemOpen
  \bibfield  {author} {\bibinfo {author} {\bibfnamefont {R.~J.}\ \bibnamefont
  {Charity}}, \bibinfo {author} {\bibfnamefont {K.~W.}\ \bibnamefont {Brown}},
  \bibinfo {author} {\bibfnamefont {J.}~\bibnamefont {Elson}}, \bibinfo
  {author} {\bibfnamefont {W.}~\bibnamefont {Reviol}}, \bibinfo {author}
  {\bibfnamefont {L.~G.}\ \bibnamefont {Sobotka}}, \bibinfo {author}
  {\bibfnamefont {W.~W.}\ \bibnamefont {Buhro}}, \bibinfo {author}
  {\bibfnamefont {Z.}~\bibnamefont {Chajecki}}, \bibinfo {author}
  {\bibfnamefont {W.~G.}\ \bibnamefont {Lynch}}, \bibinfo {author}
  {\bibfnamefont {J.}~\bibnamefont {Manfredi}}, \bibinfo {author}
  {\bibfnamefont {R.}~\bibnamefont {Shane}}, \bibinfo {author} {\bibfnamefont
  {R.~H.}\ \bibnamefont {Showalter}}, \bibinfo {author} {\bibfnamefont {M.~B.}\
  \bibnamefont {Tsang}}, \bibinfo {author} {\bibfnamefont {D.}~\bibnamefont
  {Weisshaar}}, \bibinfo {author} {\bibfnamefont {J.}~\bibnamefont
  {Winkelbauer}}, \bibinfo {author} {\bibfnamefont {S.}~\bibnamefont {Bedoor}},
  \bibinfo {author} {\bibfnamefont {D.~G.}\ \bibnamefont {McNeel}},\ and\
  \bibinfo {author} {\bibfnamefont {A.~H.}\ \bibnamefont {Wuosmaa}},\
  }\bibfield  {title} {\bibinfo {title} {Invariant-mass spectroscopy of
  $^{18}\mathrm{Ne}, ^{16}\mathrm{O}$, and $^{10}\mathrm{C}$ excited states
  formed in neutron-transfer reactions},\ }\href
  {https://doi.org/10.1103/PhysRevC.99.044304} {\bibfield  {journal} {\bibinfo
  {journal} {Phys. Rev. C}\ }\textbf {\bibinfo {volume} {99}},\ \bibinfo
  {pages} {044304} (\bibinfo {year} {2019})}\BibitemShut {NoStop}%
\bibitem [{\citenamefont {Brown}\ \emph {et~al.}(2014)\citenamefont {Brown},
  \citenamefont {Buhro}, \citenamefont {Charity}, \citenamefont {Elson},
  \citenamefont {Reviol}, \citenamefont {Sobotka}, \citenamefont {Chajecki},
  \citenamefont {Lynch}, \citenamefont {Manfredi}, \citenamefont {Shane},
  \citenamefont {Showalter}, \citenamefont {Tsang}, \citenamefont {Weisshaar},
  \citenamefont {Winkelbauer}, \citenamefont {Bedoor},\ and\ \citenamefont
  {Wuosmaa}}]{Brown:2014}%
  \BibitemOpen
  \bibfield  {author} {\bibinfo {author} {\bibfnamefont {K.~W.}\ \bibnamefont
  {Brown}}, \bibinfo {author} {\bibfnamefont {W.~W.}\ \bibnamefont {Buhro}},
  \bibinfo {author} {\bibfnamefont {R.~J.}\ \bibnamefont {Charity}}, \bibinfo
  {author} {\bibfnamefont {J.~M.}\ \bibnamefont {Elson}}, \bibinfo {author}
  {\bibfnamefont {W.}~\bibnamefont {Reviol}}, \bibinfo {author} {\bibfnamefont
  {L.~G.}\ \bibnamefont {Sobotka}}, \bibinfo {author} {\bibfnamefont
  {Z.}~\bibnamefont {Chajecki}}, \bibinfo {author} {\bibfnamefont {W.~G.}\
  \bibnamefont {Lynch}}, \bibinfo {author} {\bibfnamefont {J.}~\bibnamefont
  {Manfredi}}, \bibinfo {author} {\bibfnamefont {R.}~\bibnamefont {Shane}},
  \bibinfo {author} {\bibfnamefont {R.~H.}\ \bibnamefont {Showalter}}, \bibinfo
  {author} {\bibfnamefont {M.~B.}\ \bibnamefont {Tsang}}, \bibinfo {author}
  {\bibfnamefont {D.}~\bibnamefont {Weisshaar}}, \bibinfo {author}
  {\bibfnamefont {J.~R.}\ \bibnamefont {Winkelbauer}}, \bibinfo {author}
  {\bibfnamefont {S.}~\bibnamefont {Bedoor}},\ and\ \bibinfo {author}
  {\bibfnamefont {A.~H.}\ \bibnamefont {Wuosmaa}},\ }\bibfield  {title}
  {\bibinfo {title} {Two-proton decay from the isobaric analog state in
  $^{8}\mathrm{B}$},\ }\href {https://doi.org/10.1103/PhysRevC.90.027304}
  {\bibfield  {journal} {\bibinfo  {journal} {Phys. Rev. C}\ }\textbf {\bibinfo
  {volume} {90}},\ \bibinfo {pages} {027304} (\bibinfo {year}
  {2014})}\BibitemShut {NoStop}%
\bibitem [{\citenamefont {Egorova}\ \emph {et~al.}(2012)\citenamefont
  {Egorova}, \citenamefont {Charity}, \citenamefont {Grigorenko}, \citenamefont
  {Chajecki}, \citenamefont {Coupland}, \citenamefont {Elson}, \citenamefont
  {Ghosh}, \citenamefont {Howard}, \citenamefont {Iwasaki}, \citenamefont
  {Kilburn}, \citenamefont {Lee}, \citenamefont {Lynch}, \citenamefont
  {Manfredi}, \citenamefont {Marley}, \citenamefont {Sanetullaev},
  \citenamefont {Shane}, \citenamefont {Shetty}, \citenamefont {Sobotka},
  \citenamefont {Tsang}, \citenamefont {Winkelbauer}, \citenamefont {Wuosmaa},
  \citenamefont {Youngs},\ and\ \citenamefont {Zhukov}}]{Egorova:2012}%
  \BibitemOpen
  \bibfield  {author} {\bibinfo {author} {\bibfnamefont {I.~A.}\ \bibnamefont
  {Egorova}}, \bibinfo {author} {\bibfnamefont {R.~J.}\ \bibnamefont
  {Charity}}, \bibinfo {author} {\bibfnamefont {L.~V.}\ \bibnamefont
  {Grigorenko}}, \bibinfo {author} {\bibfnamefont {Z.}~\bibnamefont
  {Chajecki}}, \bibinfo {author} {\bibfnamefont {D.}~\bibnamefont {Coupland}},
  \bibinfo {author} {\bibfnamefont {J.~M.}\ \bibnamefont {Elson}}, \bibinfo
  {author} {\bibfnamefont {T.~K.}\ \bibnamefont {Ghosh}}, \bibinfo {author}
  {\bibfnamefont {M.~E.}\ \bibnamefont {Howard}}, \bibinfo {author}
  {\bibfnamefont {H.}~\bibnamefont {Iwasaki}}, \bibinfo {author} {\bibfnamefont
  {M.}~\bibnamefont {Kilburn}}, \bibinfo {author} {\bibfnamefont
  {J.}~\bibnamefont {Lee}}, \bibinfo {author} {\bibfnamefont {W.~G.}\
  \bibnamefont {Lynch}}, \bibinfo {author} {\bibfnamefont {J.}~\bibnamefont
  {Manfredi}}, \bibinfo {author} {\bibfnamefont {S.~T.}\ \bibnamefont
  {Marley}}, \bibinfo {author} {\bibfnamefont {A.}~\bibnamefont {Sanetullaev}},
  \bibinfo {author} {\bibfnamefont {R.}~\bibnamefont {Shane}}, \bibinfo
  {author} {\bibfnamefont {D.~V.}\ \bibnamefont {Shetty}}, \bibinfo {author}
  {\bibfnamefont {L.~G.}\ \bibnamefont {Sobotka}}, \bibinfo {author}
  {\bibfnamefont {M.~B.}\ \bibnamefont {Tsang}}, \bibinfo {author}
  {\bibfnamefont {J.}~\bibnamefont {Winkelbauer}}, \bibinfo {author}
  {\bibfnamefont {A.~H.}\ \bibnamefont {Wuosmaa}}, \bibinfo {author}
  {\bibfnamefont {M.}~\bibnamefont {Youngs}},\ and\ \bibinfo {author}
  {\bibfnamefont {M.~V.}\ \bibnamefont {Zhukov}},\ }\bibfield  {title}
  {\bibinfo {title} {Democratic decay of $^{6}\mathrm{Be}$ exposed by
  correlations},\ }\href {https://doi.org/10.1103/PhysRevLett.109.202502}
  {\bibfield  {journal} {\bibinfo  {journal} {Phys. Rev. Lett.}\ }\textbf
  {\bibinfo {volume} {109}},\ \bibinfo {pages} {202502} (\bibinfo {year}
  {2012})}\BibitemShut {NoStop}%
\bibitem [{ENS(2023)}]{ENSDF}%
  \BibitemOpen
  \href@noop {} {} (\bibinfo {year} {2023}),\ \bibinfo {note} {evaluated
  Nuclear Structure Data File (ENSDF),
  http://www.nndc.bnl.gov/ensdf/}\BibitemShut {NoStop}%
\bibitem [{\citenamefont {Cowan}(1998)}]{Cowan:1998}%
  \BibitemOpen
  \bibfield  {author} {\bibinfo {author} {\bibfnamefont {G.}~\bibnamefont
  {Cowan}},\ }\href@noop {} {\emph {\bibinfo {title} {Statistical Data
  Analysis}}}\ (\bibinfo  {publisher} {Clardendon Press},\ \bibinfo {address}
  {Oxford},\ \bibinfo {year} {1998})\BibitemShut {NoStop}%
\bibitem [{\citenamefont {Rolke}\ and\ \citenamefont
  {L\'{o}pez}(2003)}]{Rolke:2003}%
  \BibitemOpen
  \bibfield  {author} {\bibinfo {author} {\bibfnamefont {W.~A.}\ \bibnamefont
  {Rolke}}\ and\ \bibinfo {author} {\bibfnamefont {A.~M.}\ \bibnamefont
  {L\'{o}pez}},\ }\href@noop {} {\bibinfo {title} {How to claim a discovery}},\
  \bibinfo {howpublished}
  {\url{https://www.slac.stanford.edu/econf/C030908/papers/MOBT002.pdf}}
  (\bibinfo {year} {2003})\BibitemShut {NoStop}%
\bibitem [{\citenamefont {Sinervo}(2002)}]{Pekka:2002}%
  \BibitemOpen
  \bibfield  {author} {\bibinfo {author} {\bibfnamefont {P.~K.}\ \bibnamefont
  {Sinervo}},\ }\href@noop {} {\bibinfo {title} {Signal significance in
  particle physics}} (\bibinfo {year} {2002}),\ \Eprint
  {https://arxiv.org/abs/hep-ex/0208005} {arXiv:hep-ex/0208005 [hep-ex]}
  \BibitemShut {NoStop}%
\bibitem [{\citenamefont {{ATLAS Collaboration}}(2012)}]{ATLAS}%
  \BibitemOpen
  \bibfield  {author} {\bibinfo {author} {\bibnamefont {{ATLAS
  Collaboration}}},\ }\bibfield  {title} {\bibinfo {title} {Observation of a
  new particle in the search for the standard model {Higgs} boson with the
  {ATLAS} detector at the {LHC}},\ }\href
  {https://doi.org/https://doi.org/10.1016/j.physletb.2012.08.020} {\bibfield
  {journal} {\bibinfo  {journal} {Phys. Lett. B}\ }\textbf {\bibinfo {volume}
  {716}},\ \bibinfo {pages} {1} (\bibinfo {year} {2012})}\BibitemShut {NoStop}%
\bibitem [{\citenamefont {{CMS Collaboration}}(2012)}]{CMS}%
  \BibitemOpen
  \bibfield  {author} {\bibinfo {author} {\bibnamefont {{CMS Collaboration}}},\
  }\bibfield  {title} {\bibinfo {title} {Observation of a new boson at a mass
  of 125 {GeV} with the {CMS} experiment at the {LHC}},\ }\href
  {https://doi.org/https://doi.org/10.1016/j.physletb.2012.08.021} {\bibfield
  {journal} {\bibinfo  {journal} {Phys. Lett. B}\ }\textbf {\bibinfo {volume}
  {716}},\ \bibinfo {pages} {30} (\bibinfo {year} {2012})}\BibitemShut
  {NoStop}%
\bibitem [{\citenamefont {Lyons}(2013)}]{Lyons:2013}%
  \BibitemOpen
  \bibfield  {author} {\bibinfo {author} {\bibfnamefont {L.}~\bibnamefont
  {Lyons}},\ }\href@noop {} {\bibinfo {title} {Discovering the significance of
  5$\sigma$}} (\bibinfo {year} {2013}),\ \Eprint
  {https://arxiv.org/abs/1310.1284} {arXiv:1310.1284 [physics.data-an]}
  \BibitemShut {NoStop}%
\end{thebibliography}
%

\end{document}